%Paper: astro-ph/9511005
%From: Basilio Santiago <santiago@ast.cam.ac.uk>
%Date: Wed, 1 Nov 1995 16:44:50 +0000 (GMT)
%Date (revised): Thu, 2 Nov 1995 18:35:50 +0000 (GMT)

\magnification=\magstep1
\hoffset=-0.6 true cm
\voffset=0.0 true cm
\baselineskip=20pt minus 1pt
\baselineskip=14pt

\vsize=8.9truein
\hsize=6.8truein
%\hyphenpenalty=10000
%\raggedright
%\parskip=\medskipamount
\tolerance=10000
\parindent=1truecm
\raggedbottom
\def\pp{\parshape 2 0truecm 5.8truein 2truecm 5.01truein}
\def\ltsima{$\; \buildrel < \over \sim \;$}
\def\simlt{\lower.5ex\hbox{\ltsima}}
\def\gtsima{$\; \buildrel > \over \sim \;$}
\def\simgt{\lower.5ex\hbox{\gtsima}}
\def\bline{\hbox to 1 in{\hrulefill}}
\def\etal{{\sl et al.\ }}
\def\np{\vfill\eject}

\def\iras{{\sl IRAS\/}}
\def\dof{{\rm dof}}
\def\bfr{{\bf r}}
\def\kms{\ifmmode {\rm \ km \ s^{-1}}
\else
$\rm km \ s^{-1}$\fi}

\def\vp{\ifmmode {\bf r}
\else
${\bf r}$\fi}
\def\self{\ifmmode {\phi ({\rm r})}
\else
{$\phi ({r})$}\fi}
\def\sfvp{\ifmmode {\phi ({\rm {\bf r}})}
\else
$\phi ({{\bf r}})$\fi}
\def\lumf{\ifmmode {\Phi ({\rm L})}
\else
{$\Phi ({L})$}\fi}
\def\diamf{\ifmmode {\Phi ({\rm D})}
\else
{$\Phi ({D})$}\fi}
\def\lfvp{\ifmmode {\Phi ({\rm L,{\bf r}})}
\else
{$\Phi ({L,{\bf r}})~$}\fi}
\def\n1{\ifmmode {{n_1}}
\else
${{n_1}}$\fi}
\def\ab{\ifmmode {A_B}
\else
${A_B}$\fi}
\def\ablb{\ifmmode {A_B (l,b)}
\else
${A_B (l,b)}$\fi}
\def\mlim{\ifmmode {\rm m_{lim}~}
\else
${m_{lim}~}$\fi}
%\nopagenumbers
%\headline{\hfill \folio \hfill}

\centerline{\bf THE OPTICAL REDSHIFT SURVEY II:}
\medskip
\centerline{\bf DERIVATION OF THE LUMINOSITY AND DIAMETER FUNCTIONS}
\centerline{\bf AND OF THE DENSITY FIELD\footnote{$^1$}{\rm Based in
part on data obtained at Lick Observatory, operated by
the University of California; the Multiple Mirror Telescope, a joint
facility of the Smithsonian Astrophysical Observatory and the
University of Arizona; Cerro Tololo Inter-American Observatory,
operated by the Association of Universities for Research in Astronomy,
Inc., under contract with the National Science Foundation; Palomar
Observatory, operated by the California Institute of Technology, the
Observatories of the Carnegie Institution and Cornell University;
and Las Campanas Observatory, operated by the
Observatories of the Carnegie Institution.\vskip 1pt}}
\bigskip
\centerline{Bas\'\i lio X. Santiago\footnote{$^2$} {Institute of Astronomy,
Cambridge University, Madingley Road, Cambridge CB3 0HA, United
Kingdom\vskip 1pt},
Michael A. Strauss\footnote{$^3$} {Institute for Advanced Study,
School of Natural Sciences, Princeton, NJ 08540 U.S.A.\vskip 1pt}, Ofer
Lahav$^2$, }
\centerline{Marc Davis\footnote{$^4$} {Physics and Astronomy Departments,
University of California, Berkeley, CA 94720 U.S.A.\vskip 1pt}, Alan
Dressler\footnote{$^5$}{Observatories of the Carnegie Institution of
Washington, 813
Santa Barbara St., Pasadena, CA 91101 U.S.A.\vskip 1pt}, and John P.
Huchra\footnote{$^6$}
{Center for Astrophysics, 60 Garden St., Cambridge, MA 02138 U.S.A.}}
\bigskip
\bigskip
\centerline{\it Submitted to The Astrophysical Journal}
\np
\centerline{\bf ABSTRACT}
\medskip
We quantify the effects of Galactic extinction
on the derived luminosity and diameter functions and on the
density field of a redshift sample.
Galaxy magnitudes are more affected by extinction than are diameters,
although the effect on the latter is more variable from galaxy to
galaxy, making it more difficult to quantify.  We develop a
maximum-likelihood approach to correct the luminosity function, the
diameter function and the density field for extinction
effects. The effects of random and systematic photometric errors
are also investigated.
The derived density field is robust to both random and systematic
magnitude errors as long as these are uncorrelated with position on the sky,
since biases in the derived selection function and number counts tend
to cancel one another.
Extinction-corrected luminosity and diameter functions are
derived for several subsamples of the {\it Optical Redshift Survey} (ORS).
Extinction corrections for the diameter-limited subsamples are found to
be unreliable, possibly due to the superposition of random and systematic
errors. The ORS subsamples are combined
using overall density scaling factors from a full-sky redshift survey of
\iras\ galaxies, allowing the reconstruction of the optical
galaxy density field over most of the sky to 8000 \kms.
\bigskip
\centerline {\bf 1. Introduction}
\medskip
Santiago \etal (1995; hereafter Paper I) presented the {\it Optical
Redshift Survey} (ORS). The ORS contains $\sim 8500$ galaxies with
almost complete redshift information, selected from the UGC (Nilson
1973), ESO (Lauberts 1982; Lauberts \& Valentijn 1989), and ESGC
(Corwin \& Skiff 1995) catalogues. The sample covers most of the sky
with $|b| > 20^\circ$ in both hemispheres, and consists of two
largely overlapping subsamples, limited in apparent magnitude and
diameter, respectively. This sample is used to
derive the galaxy density field to $cz
\approx 8000$ \kms\ and, in future papers in this series, to  study its
dependence on galaxy
morphology and spectral properties. Its dense sampling and large
variety in galaxy content make the ORS ideal for this
purpose. Nevertheless, since there is no single optical galaxy
catalogue covering the entire sky that probes significantly beyond the
Local Supercluster, the ORS sample selection is based on three different
catalogues and thus is not uniform over the sky.
There may also be systematic trends in the magnitude and
diameter scales internal to the different galaxy catalogues from which
ORS was derived.  In addition, the ORS extends to low Galactic
latitudes, where variable
Galactic absorption introduces an additional
source of systematic variations in the sample selection. Previous
studies of the large-scale distribution of galaxies have either worked
in passbands where Galactic extinction was unimportant (e.g., Strauss
\etal 1990) or have restricted themselves to regions of the sky in which
the measured extinction was low (e.g., da Costa \etal 1994; cf.,
Appendix A of Strauss \etal 1992b).
In this paper, our aim is to measure the density field of galaxies in
the face of extinction and photometric errors; this paper describes
our techniques for doing so, and our tests of these methods.

Recovering the galaxy density field requires determining the correct
selection function for the sample. The selection function quantifies
the loss of galaxies due to the magnitude or diameter limit and to any other
selection effect that may be present.
Thus, non-uniformities in survey sampling must be
correctly incorporated by the derived selection function.
The selection function is also intimately related to the galaxy luminosity
and diameter functions in magnitude or diameter-limited
surveys, respectively.  The latter are cosmologically relevant
quantities; their shape and dependence on environment density or on
galaxy intrinsic properties have the potential to constrain
different scenarios of
structure formation (Blumenthal \etal 1984; White \& Frenk 1991; Cole
\etal 1994).  A precise determination of the luminosity function and diameter
function at $z = 0$ is
necessary when studying the evolution of these functions or when comparing
observed galaxy number counts with model predictions
(Davis, Geller \&
Huchra 1978, Dekel \& Rees 1987, Koo \& Kron 1992).
Previous determinations of the luminosity function of optical galaxies
have been carried out by
Kirshner, Oemler, \& Schechter (1979), Sandage,
Tammann \& Yahil (1979), Davis \&
Huchra (1982), Kirshner \etal (1983), Efstathiou, Ellis, \& Peterson
(1988), de Lapparent, Geller, \& Huchra (1989), Santiago \& Strauss
(1992), Loveday \etal (1992, 1995), and Marzke, Geller, \& Huchra
(1994), among others.
Different parametric and non-parametric methods for determining the
luminosity function and diameter function have been developed and
applied; these methods are unbiased by the presence of inhomogeneities in the
galaxy
distribution (Turner 1979, Sandage \etal 1979, Nicoll \& Segal 1983,
Choloniewski 1986, Efstathiou \etal 1988; see Binggeli, Sandage, \&
Tammann 1988 for a review). The diameter function has been
investigated by Maia \& da Costa (1990), Lahav, Rowan-Robinson, \&
Lynden-Bell (1988) and Hudson \& Lynden-Bell (1991). Finally, the
bivariate distribution of diameters and absolute magnitudes has been
analyzed by Choloniewski (1985) and Sodr\'e \& Lahav (1993).

In this paper we discuss the effects of Galactic absorption and
systematic and random magnitude or diameter errors on the derived
selection function and density field. In \S 2 we quantify the
extinction effects by means of Monte-Carlo simulations. A
maximum-likelihood method is derived to recover an unbiased estimate
of the selection function and of the density field. The influence of
other systematic and random errors is also addressed. In \S 3, we
apply the methods developed in \S 2 in order to recover the ORS
luminosity and diameter functions.  Our main conclusions are
presented in \S 4.  Throughout this paper, we work in redshift space;
thus we are not concerned here about corrections from redshift to real
space for galaxies in redshift surveys.
\bigskip
\centerline {\bf 2. Extinction and Systematic and Random Errors}
\medskip\nobreak
\centerline {\it 2.1 The Selection Function and Galactic Extinction}
\medskip\nobreak
Any magnitude or diameter-limited sample will become sparser at larger
redshifts due to the increasing loss of galaxies caused by the adopted
apparent magnitude or diameter cut-off.  This effect is quantified by
the selection function, which expresses the fraction of the total
population of galaxies that are expected to satisfy the sample's
selection criterion.  In the case of a flux-limited sample, the
selection function is given by
$$\phi (r) = {\int_{L_{min} (r)}^\infty
\Phi(L) dL \over \int_{L_s}^\infty \Phi(L) dL}, \eqno (1)$$
where $L_{min} (r) \equiv 4\pi r^2 f_{min}$ is the minimum luminosity
necessary for a galaxy at distance $r$ to make into the sample,
$f_{min}$ is the limiting flux\footnote{$^7$}{In the case of a survey
limited in magnitudes to $m_{lim}$, $f_{min} \propto 10^{-0.4
m_{lim}}$.} of the sample in erg$\rm\,s^{-1}\,cm^{-2}$,
and $\Phi (L)$ is the luminosity function in the passband
in which the survey is done. In principle, the lower limit of the
integral in the denominator should be zero, but we set it to a small
value $L_s \equiv L_{min}(r_s)$, as it is impossible in practice to
quantify the faintest end of the luminosity function. We set $r_s =
500 \kms$ in what follows.  In the equation above and throughout this
paper we assume that $\Phi (L)$ is universal, showing no dependence on
location in the Universe. We will test this assumption {\it a
posteriori\/} in \S 3.

Yahil \etal (1991; hereafter YSDH) adopt a simple parameterized form
for the selection function, characterized by three numbers, which
experience has shown is sufficiently
general for these purposes (e.g., Santiago \& Strauss 1992):
$$
\phi(r) = \cases{\left({r \over {r_s}}\right)^{-2 \alpha} \left ({{ r_*^2 +
r^2} \over {r_*^2 + r_s^2}}\right)^{-\beta}& $r > r_s$\cr
1 & $r \le r_s$\cr}\quad . \eqno(2)
$$
Equation (1) implies that the luminosity function can be related to
the selection function by a derivative, yielding:
$$\Phi (L) = C \left( {\alpha \over L} + {\beta \over {L_* + L}}
\right) \left( {L \over L_*} \right)^{-\alpha} \left( 1 +
{L \over L_*} \right)^{-\beta}, \eqno (3)$$
where $\alpha$, $\beta$
and $L_*\equiv 4\,\pi r_*^2 f_{min}$ are free parameters and $C$ is a
normalization constant. Thus $r_*$ is the distance at which galaxies
at the survey flux limit have a luminosity characteristic of that of
the knee of the luminosity function. The best-fit values for these
parameters are given below in \S 3. This form is a generalization of the
Schechter
(1976) function with one additional free parameter (cf., Yahil
1988). We will use the parameterization of Equation (2) in this paper,
but Equation (1) will have to be generalized to reflect the effects of
extinction. Before doing so, let us remind ourselves how one would
solve for the selection function for the sample in the case of no
extinction. Following YSDH, we
maximize the likelihood $\Lambda$, the product of probabilities $p$
 that each galaxy have its measured
luminosity, given its redshift. For a purely flux-limited sample,
without the effects of extinction, this is:
$$ \eqalignno
{\Lambda &= \prod_{i}^N p(L_i \mid r_i) \cr
         &= \prod_{i}^N {\Phi (L_i) \over
\int_{L_{min} (r_i)}^\infty \Phi (L^{\prime}) dL^{\prime}}\cr
         &\propto \prod_{i}^N {1 \over \phi (r_i)}
{d\phi (r) \over dr} \Bigg\vert_{r_{max} (L_i)}\quad,&(4)\cr}$$
where $N$ is the total number of galaxies in the sample and $r_{max} \equiv
r (f/f_{min})^{1/2}$ is the maximum distance out to which the galaxy
could be placed and remain in the sample. The last step follows from
Equation (1).

The derived selection function is used for recovering the
density field; the sampling selection effect is compensated by
weighting each galaxy by the inverse of the selection function at
its position, $w_i = 1/\phi(r_i), i=1, \ldots, N$. Densities can then be
computed
by summing each galaxy's contribution within a given volume
and dividing by that volume. We usually express densities in units of
the global mean density
$$n_1 = \sum w_i/V,\eqno(5)$$
 computed over the entire
sample volume $V$ out to some limiting redshift (see YSDH, Eqns.~13-15).

Now let us consider the effects of extinction.
Consider a galaxy whose intrinsic apparent photographic $B$ magnitude
is given by $m$.  If the galaxy lies in the direction $(l,b)$, its
observed magnitude is given by
$$ m_{obs} = m + \gamma_m A_B(l,b), \eqno (6) $$
where \ablb\ is the extinction in the given direction and
passband. If $m_{obs}$ is the total observed magnitude, $\gamma_m =
1$; extinction simply reduces the surface brightness at each point on
the galaxy's two dimensional figure by \ablb.  For isophotal
magnitudes, however, not only does every point in the galaxy become
dimmer, but the radius of the isophote shrinks, and thus we expect
$\gamma_m > 1$, the exact value depending on the two-dimensional light
profile of the object (Cameron 1990).
For example, the effects of extinction for a spiral disk with an
exponential light profile of $e$-folding length $\theta_e$ can be
calculated analytically. The observed isophotal diameter $\theta_{obs}$
is related to the intrinsic isophotal diameter $\theta$ by:
$$\theta_{obs} = \theta - 0.921 \,\theta_e A_B. \eqno(7)$$
The quantity $\gamma_m$ as defined in equation (5) is then
$$\gamma_m =  1 + {2.5 \over \ab }~\log { {1 - e^{-x} (1 + x)} \over
{1 - e^{-x_{obs}}(1 + x_{obs})}} \approx {{1 - e^{-x}}\over{1 - e^{-x} (1
+ x)} }\eqno(8)
$$
where $x \equiv \theta/\theta_e$ and  $x_{obs} \equiv
\theta_{obs}/\theta_e$ are the number of scale lengths subtended by
the intrinsic and observed diameters, respectively. The second
equality holds in the limit of small $A_B$.
Notice that
the second term on the right-hand side of Equation~(8) depends
on the profile shape through the number of $e$-folding lengths
within $\theta$ and $\theta_{obs}$.

We can similarly define a quantity $\gamma_d$ which gives the
fractional decrease in isophotal diameter with extinction:
$$\gamma_d \equiv {{5} \over {A_B}} \log {{\theta} \over{\theta_{obs}}}
\approx {{2.0}\over x},\eqno(9)$$
where again, the second equality holds in the limit of small $A_B$.

For a given surface brightness profile, the quantities $\gamma_m$ and
$\gamma_d$ depend only on  $x$ and
\ab. They are related to the fractional changes in flux and diameters
by the equations $\delta f/f = 1 - 10^{-0.4 A_B \gamma_m}$ and $\delta
\theta/\theta = 1 - 10^{-0.2 A_B \gamma_d}$, respectively.  Figure~1
shows $\gamma_d$ and $\gamma_m$ as a function of $\theta / \theta_e$
for $A_B = 0.3$ mag; the dependence on $A_B$ is weak for small and
moderate \ab.  Panel {\it a} assumes an exponential disk and panel
{\it b} refers to a de Vaucouleurs profile, for which expressions
analogous to Equations (7)-(9) can be derived.  In the latter case,
$\gamma_m$ also depends weakly on $\theta_e$; we assumed a typical
value for
early-type galaxies.  In both panels, the horizontal bars indicate the
observed range of $\theta / \theta_e$ values for disk and bulge
galaxies, assuming limiting isophotal levels for detection
characteristic of the ESO and Palomar plates. Note that $\gamma_m$ is
significantly larger than $\gamma_d$ for any value of
$\theta/\theta_e$; fluxes are more
affected by extinction than are diameters. However, $\gamma_d$ is a
stronger function of $x$ than is $\gamma_m$, since $\gamma_d$ approaches
0 for large extinction, while $\gamma_m$ approaches 1. Thus the
approximation we will use of a single value of $\gamma$ for an entire
sample is likely to be less appropriate for diameter-limited samples
than for magnitude-limited samples.

For a flux-limited sample, the effect
of Galactic obscuration is to introduce a non-uniform, direction-dependent
selection effect: the effective
magnitude cut-off limit for a galaxy lying in the direction $(l,b)$
is given by
$m_{lim} = m_{lim,\,observed} - \gamma_m \ablb$, where
$m_{lim,\,observed} \equiv {\rm const} - 2.5 \log f_{min}$ is the magnitude
corresponding to the flux limit.
In terms of the selection
function, we now have
$$\phi_{obs} [\bfr,\gamma_m A_B(l,b)] =
{\int_{L_{min} (r,\gamma_m A_B)}^\infty \Phi(L) dL
\over \int_{L_s}^\infty \Phi(L) dL}, \eqno (10)$$
where
$L_{min} (r,\gamma_m A_B) \equiv 4 \pi r^2 f_{min} 10^{0.4 \gamma_m \ab}$ is
the
minimum luminosity necessary for a galaxy in this direction to
make it into the sample. A similar expression holds for diameter-limited
samples.

In the absence of absorption ($\ablb = 0$), Equation
(10) becomes the usual definition of the selection function,
\self\ (Equation 1),
showing no dependence on direction in the sky.
We call this quantity the intrinsic selection function $\phi(r)$, in order
to distinguish it from the selection function that incorporates
extinction, $\phi_{obs} (\bfr,\gamma A_B)$.
The two functions are related by the equation
$$\phi_{obs} (\bfr,\gamma \ab) = \phi (r 10^{0.2 \gamma \ab}). \eqno (11) $$

Notice that it is $\phi$ rather than  $\phi_{obs}$ that
is directly related to $\Phi (L)$ via Equation (1). On the other hand,
it is $\phi_{obs}$ that quantifies the extinction selection effects
inherent to the sample. We thus need \self\ to derive the luminosity
function, and $\phi_{obs}$ to derive an unbiased density field.

What are the errors resulting from not taking Galactic absorption into
account when deriving the selection function and density field?  We
answer this question by means of simulations of a redshift survey.  We
use a standard Cold Dark Matter (CDM) N-body simulation ($\Omega = 1, h=0.5,
\Lambda = 0$) generated by OL and Karl Fisher, and select a random
subsample of points around an origin chosen to represent the observer,
according to an input selection function. Our results on the effect of
extinction do not depend on the details of the cosmological model or
selection function adopted.  Fluxes are assigned to each selected
point following the luminosity function consistent with the chosen
intrinsic selection function.  A ``Galactic'' coordinate system is set
up as well, and each magnitude is attenuated according to the
Burstein-Heiles (1982) extinction in that direction.  The value of
$\gamma$ used for each point is drawn from a Gaussian distribution
with mean unity and dispersion $\sigma = 0.2$. We thus allow for
variations in the extinction correction from object to object, and
uncertainties in the extinction map.  The sample was then cut at
$m_{obs} \leq 14.5$. At this point, we wish to search for biases in
our techniques for correcting for extinction, as opposed to assessing
random errors in the derived quantities, and thus we generated a random
catalogue with over 10,000 points, roughly twice that of the ORS sample
itself.

We first compute the selection function
for the sample without attempting to apply any correction
for extinction effects; we make straight use of the
method described in YSDH and outlined above.
The input \self\ used in this and all subsequent Monte-Carlo
experiments in this paper is shown in Figure~2a.
The ratio of the derived to the input  selection function
is shown in Figure~2b as a dashed line.
The monotonic decrease in the ratio is caused by the loss of galaxies
{}from the sample due to Galactic absorption.
However, this discrepancy is due mainly to the mean extinction over
the sample; it is not sensitive to the variation of the extinction
{}from point to point. The dotted line in Figure~2b is the quantity
$\phi_{obs}/\self$, where $\phi_{obs}$ was
calculated from Equation (11), using the input selection
function and the mean $A_B$ taken over all the points in the
simulation. This scaled selection function is very similar to the
one derived directly from the sample.

It is the intrinsic selection function \self\ that we are interested
in recovering. We can generalize the
maximum-likelihood expressions of Equation
(4) to take extinction into account:
$$ \eqalignno
{\Lambda &= \prod_{i}^N p(L_i \mid r_i,\gamma_i A_B^i) \cr
         &= \prod_{i}^N {\Phi (L_i 10^{0.4 \gamma_i A_B^i}) \over
\int_{L_{min} (r_i,\gamma_i A_B^i)}^\infty \Phi (L^{\prime}) dL^{\prime}}\cr
         &= \prod_{i}^N {1 \over \phi (10^{0.2 \gamma_i A_B^i} r_i)}
{d\phi (r) \over dr} \Bigg\vert_{{\rm dex}({0.2 \gamma_i A_B^i})
r_{max} (L_i)}\quad ,&(12)\cr}$$
where $N$ is the total number of galaxies in the sample.

Equation (12) takes extinction directly into account when fitting for
the parameters of \self.
However, we do not have {\it a priori\/} knowledge of the values of $\gamma_i,
i=1,\ldots,N$. This would require knowing the surface brightness profiles of
all galaxies in the sample.
We thus make the approximation of a single value of $\gamma$ for all
galaxies, as if they all had the same surface brightness
profile. $\gamma$ is then treated as an additional parameter to be
fitted simultaneously to the luminosity (or selection) function. In
Figure~2c we show the result of maximizing this new likelihood
function for the same Monte-Carlo simulation as in Figure~2b. The
derived selection function is now in very good agreement with the
input one.  Given the intrinsic selection
function, $\phi_{obs} (r,\gamma A_B)$ for each galaxy follows directly from
Equation (11). Thus this approach allows us to successfully recover the
correct luminosity function (free of extinction biases) for a redshift
sample, even in the presence of moderate variations in the true value
of $\gamma$ from galaxy to galaxy.

We next tested our ability to measure $\gamma$ itself. We carried out
a series of simulations in which the true mean value of $\gamma$ was
varied; in each case, the value of $\gamma$ for each galaxy followed from a
Gaussian distribution with a dispersion of 0.2. Figure~3a shows the
derived $\gamma$ values as a function of the mean input $\gamma$.
The solid points were derived from simulations of size similar to that
used in Figure~2, containing about 10,000 points.
The open symbols were obtained from simulations whose sizes are
comparable to the ORS subsamples used in \S 3 and that had
random magnitude errors (typically of 0.5 mag) added to the extinction effect.
Formal error bars from the maximum-likelihood fits are plotted; the
smaller simulations have larger error bars, as expected.
The agreement between derived and input $\gamma$ values is
still good in all cases, although there is a small trend
towards underestimating
large input $\gamma$ for the small sample.
We investigated the origin of this trend by plotting the derived
selection functions (normalized to the input one)
for each ORS-sized simulation in
Figure~3b. Contrary to the error free case, \self\ is now shallower
than the input function used (ratio greater than unity).
This is caused by Malmquist bias, as will be explained
in \S 2.3. Notice, however, that the amplitude
of the bias in \self\ is insensitive to $\gamma$,
indicating that the trend in $\gamma$ seen in Figure~3a does not
lead to discrepant best-fit solutions for \self. Instead, the
systematics in $\gamma$ just
compensate for similar biases in the \self\ parameters; indeed, the
derived density field (using the methods discussed in connection with
Figure~5a below) shows no systematic bias that grows with $\gamma$.
We thus conclude that our revised likelihood
approach recovers the right \self\ (apart from Malmquist bias)
even when $\gamma$ is significantly different from unity and
in the presence of random scatter in the magnitudes.

We now consider the effect Galactic absorption has on the derived
density field. We again use the Monte-Carlo simulation with input mean
$\gamma = 1$.  Galaxies are weighted by the inverse of the derived
selection function, $\phi_{obs} (r,\gamma A_B)$, and smoothed with a tophat
of radius given by the mean interparticle separation at each
point. Densities $\rho(\vp)$ are measured on a grid of points in space,
interspaced by the mean interparticle distance at their location.  We
normalize the densities by $n_1$ as given by Equation (5), the global
mean computed over the whole sample volume to 8000 \kms, and define
deviations from the mean as $\delta(\vp) \equiv \rho(\vp)/\langle \rho
\rangle - 1$.

Figure~4 shows how these grid point normalized densities, $\delta(\vp)$,
are affected by Galactic absorption.  In panel {\it a} we plot the
errors in the density, $\Delta \delta(\vp) \equiv \delta(\vp) -
\delta_{obs}(\vp)$, as a function of \ab.
The quantity $\delta_{obs} (\vp)$
was computed by weighting each galaxy by the inverse of the selection
function which completely ignores extinction (Figure~2b). We show the
mean value of $\Delta \delta$ for different bins in $A_B$ (solid line).
There is clearly a trend in the sense that densities are
underestimated (overestimated) in regions of high (low) absorption.
This is a direct consequence of the insensitivity of the selection
function to variations in extinction over the sky.  In panel {\it b},
$\Delta \delta$ is plotted as a function of redshift. The mean over all
points in each distance bin is again plotted as a solid line. Also
plotted are the mean curves for the points in each bin with the 25\%
highest and lowest \ab\ values.  Notice that $\Delta \delta$ for the
upper quartile tends to increase with distance, whereas the lower
quartile shows a systematic decrease: the bias in the
density field caused by extinction increases with distance. This is
consistent with the increasingly larger disagreement between the
derived and input selection functions in Figure~2b. At $v \sim 4000
\kms$, the error in density can be as large as $\Delta \delta
\sim 1.0$, although the typical values are around 0.2. The results
shown are for $\langle \gamma \rangle = 1$; these effects become
larger for larger values of $\gamma$.
On the other hand, there is no systematic bias in the mean density at
a given distance; at each distance, there
will be directions with both high and low $A_B$ values, whose
densities are underestimated and overestimated, respectively. We thus
conclude that ignoring extinction tends to cause biases only in local
density estimates, while densities averaged over the solid angle of a
survey are not systematically
affected.

Finally, in panel {\it
c}, we show how the difference between intrinsic and derived densities
behaves as a function of the intrinsic density itself.  Again, no
systematic trend is seen when all points are considered.  However, for
the high and low \ab\ quartiles, we see errors of about
20-25\% the mean density.

Figure~5 shows the results of weighting each galaxy by the inverse of
$\phi_{obs} (r,\gamma A_B)$ as derived by simultaneously fitting \self\
and $\gamma$ and then applying Equation (11) to each object.  No
systematic error in the densities is now visible either as a function
of $A_B$, distance or intrinsic density. In particular, the mean
curves for the highest and lowest extinction quartiles (panels {\it b}
and {\it c}) are both very close to zero. In addition, the scatter
in the diagrams has
decreased substantially. The remaining scatter is almost entirely due to
shot-noise; rederiving the density field for the Monte-Carlo sample
used in Figure~5, but with exact input values of the $\gamma_i$ and
of the selection function made almost no difference in the scatter.

In summary, we have found that the bias in the density field caused
by extinction is a direct consequence of that in the selection
function. In particular, by trying to fit a selection function which
depends only on distance
to a sample affected by Galactic absorption, we do not incorporate the
variability of the extinction as a function of direction. Local density
estimates are then systematically in error but densities averaged over
solid angle at a given distance are not
affected. By taking extinction explicitly into account when deriving
the selection function we recover both the
correct galaxy luminosity function and density field, even in the
presence of scatter in the value of $\gamma$ (as long as that scatter
is uncorrelated with position on the sky). Most of the remaining
scatter present in the estimated densities is the result of shot-noise.
\bigskip
\centerline {\it 2.2 Other Systematic Effects}
\medskip\nobreak
As far as their influence on the derived density field is concerned,
systematic errors fall into two categories: position-dependent and
position-independent errors.
In the former case, without {\it a priori\/} information on
the form of the error (such as we have in the form of the
Burstein-Heiles maps in the case of extinction), one cannot solve or
test for it. In the latter case, however, a fit to a radial selection function
will absorb the systematic effect\footnote{$^8$}{This is true to the
extent that the functional form of the selection function to which we
are fitting is sufficiently general to be able to absorb the
systematic effect.}, leaving the derived density unbiased.

Consider, for
instance, the case in which the observed magnitudes are incorrect by
both an offset and a scale error (such as might be caused by
uncorrected non-linearities in photographic plate material):
$$m_{obs} = a m + b, \eqno (13)$$
where $a$ and $b$ are constants and
$m$ represents the true magnitude. The selection function that
accommodates this kind of magnitude error is given by Equation
(1) with $L_{min} (r) = 4 \pi r^2 f_{min}^{1/a} 10^{0.4 b/a}$.
Notice that both the selection function and the cut-off luminosity
$L_{min}$ are still functions of distance only.  Our maximum
likelihood method will thus find the correct selection function even
in the presence of such systematic effects, as we demonstrate with a
Monte-Carlo simulation.
Observed magnitudes were obtained from Equation (13) with $a = 1.04$,
$b = -0.3$, giving a systematic effect comparable to that of
extinction. This simulation did not include the effects of extinction
or random magnitude errors.
The trend given by Equation (13) is assumed unknown {\it
a priori\/}, and as such is not explicitly corrected for when deriving
\self.  Figure~6a shows the ratio of derived to input selection functions.
The former is steeper than the latter due to the loss of galaxies
caused by the systematic trend in the magnitudes.  The corresponding
\lumf\ will be then equally biased.  The derived density field,
however, shows only a small bias, both as a function of distance
(panel {\it b}) and intrinsic density (panel {\it c}). The residual
bias increases if we make the systematic error given by Equation (13)
stronger, since the adopted parameterization for \self\ becomes increasingly
inadequate.

On the other hand, consider an error in which the magnitudes in the
Northern Hemisphere are systematically brighter than are those in the
South, by an amount $\Delta_0$.
If we do not know the value of $\Delta_0$ but are aware of a
systematic offset between the two hemispheres, we can
account for the resulting variations in sampling by
computing the selection function separately for the North and South.  By
restricting the maximum-likelihood fit to galaxies which are subject
to the same value of $\Delta$, we are in practice determining
$\phi_{obs} (\vp) =
\phi (r 10^{0.2 \Delta})$.
The density field calculated by combining the two
hemispheres, each with their own selection function, will not be biased
by sample inhomogeneity. However, {\it
if the mean density within each hemisphere is not equal to the global mean
density} the normalization applied to the derived densities will be wrong.
This means that a sample with systematic errors between
different hemispheres cannot be used to calculate such quantities as
the dipole moment of the galaxy distribution (e.g., Strauss \etal
1992c). In order to properly normalize the {\it global\/} density
field in such a
case, one needs external information on the density field, such as
supplied by the \iras\ redshift survey (Strauss \etal 1992a; Fisher
\etal 1995). This is the approach we use in \S 3 below.
\bigskip
\centerline {\it 2.3 Random Errors}
\medskip\nobreak
Random errors in the magnitudes have two effects on the derived
density field in a magnitude-limited sample: first, galaxies scatter
across the magnitude limit $m_{lim}$. Because galaxies with true
magnitudes fainter than
$m_{lim}$ outnumber those brighter than $m_{lim}$, the net effect is
to augment the numbers of galaxies over the case of no errors, an
effect which increases with distance. Second, the magnitude errors
convolve directly with the luminosity function to broaden it, thus
making the selection function shallower. These two effects work in
opposite senses, and thus we might hope that they largely cancel one
another in the density field.
Let the distribution of errors $\epsilon$ in the fluxes be
$g(\epsilon, f)$, where the distribution
can depend on $f$ itself. Suppose the density fluctuation at a point
\bfr\ is $\delta(\bfr)$; the number of galaxies expected within a
neighborhood $d V$ of \bfr\ is\footnote{$^9$}{We have not included the
effects of extinction, for simplicity.}:
$$
N_{obs} (\bfr) = [1 + \delta(\bfr)]\,n_1 d V \int_{L'_{min}(r)}^\infty d L'
\int_{-\infty}^\infty g(\epsilon,f) \Phi[4 \,\pi r^2 f] d
\epsilon;\eqno(14)$$
$L' \equiv 4\,\pi r^2 (f + \epsilon)$ is the observed luminosity, and $4
\,\pi r^2 f$ is the true luminosity. The observed
selection function $\phi_{obs}(\bfr)$ is given by an integral over
observed luminosities, down to $L'_{min}(r)$, of the observed
luminosity function. The observed luminosity function is the
convolution of the true luminosity function with the error
distribution function, thus
$$
n_1 \phi_{obs}(r) = \int_{L'_{min}(r)}^\infty d L'
\int_{-\infty}^\infty g(\epsilon,f) \Phi[4 \,\pi r^2 f] d
\epsilon,\eqno(15)$$
which is the same double integral as in Equation~(14). Thus the
observed density field is given by
$$1 + \delta_{obs}(\bfr) \equiv {{N_{obs}(\bfr)}\over{n_1 \phi_{obs}(\bfr) d
V}} = 1 + \delta(\bfr).\eqno(16)$$
Therefore {\it the density field is unbiased by random errors in the
fluxes.} The one assumption we made was that the error distribution
was independent of position in the sky.

We demonstrate this with Monte-Carlo simulations. We assume a Gaussian
error distribution in magnitudes with zero mean. In Figure~7a, we plot
the derived selection functions for two different choices of Gaussian
width $\sigma$. We also plot the result for the case of $\sigma$
depending on magnitude: $\sigma = 0.2 $ for $m \leq 13$ and increasing
linearly with $m$ to $\sigma = 0.5$ for $m \geq 14.5$. The selection
function becomes increasingly flatter (ratio increasing) with
increasing $\sigma$, due to Malmquist bias.  In panels {\it b} and
{\it c} of Figure~7, respectively, we show the mean difference between
intrinsic and derived densities for each case, as a function of
density and distance.  No systematic differences are found as a
function of distance. There is a small tendency for underdense regions
to have their densities overestimated, and overdense regions to have
their densities underestimated due to the effect of random errors, but
this is at most a 10\% effect for the examples shown. This confirms
the robustness of the density field to random errors in the
magnitudes. Finally, panel {\it d} shows the rms error in the density
field as a function of the density; although the bias does not grow
with increasing $\sigma$, the error certainly does.

\bigskip
\centerline {\bf 3. ORS Results}
\medskip\nobreak
In this section we derive luminosity and diameter functions
for ORS galaxies. We also discuss the amplitude of extinction
effects on the data and its dependence on morphological types.
Finally, we show the density field projected on
several different slices of the local Universe.

The three catalogues from which the ORS sample was defined
are based on separate plate material and were done by different
observers. In anticipation of differences in the various photometric
systems, we derived the selection
functions separately for each catalogue, for both magnitude and
diameter-limited
samples.
Luminosity and diameter functions then follow automatically
{}from the inferred \self. As we saw in the previous sections,
these are probably biased by unknown systematic and random errors
internal to each catalogue from which ORS was extracted.
Systematic errors within each catalogue's limits are difficult to
quantify. There are indications, for instance, that the magnitudes
{}from Volume I of the CGCG (Zwicky \etal 1961-1968) are
systematically fainter by $\Delta = 0.3$ mag
than those from the other volumes (Kron \& Shane 1976, Paturel 1977).
Instead of computing a separate selection function for this volume,
we explicitly correct its magnitudes just as we do for
extinction.
As for random errors, they are estimated to be of
approximately $\Delta m \simeq
0.09$ and $\Delta m \simeq 0.3$ for ESO and Zwicky magnitudes,
respectively (Lauberts \&
Valentijn 1989, Kron \& Shane 1976, Huchra 1976).
Errors in the diameters are typically $\delta \log D$ \ltsima 0.09
(Paturel \etal 1987, 1991; de Vaucouleurs \etal 1991).
According to the results shown in Figure~7, the associated biases in
the density field are typically not larger than 10\% of the mean density.
The quoted errors, however, may depend on such parameters
as inclination, morphological type or magnitudes and diameters themselves.
Since our main interest is in the density field, which we found to be
less vulnerable to random and systematic errors than is the luminosity
function, we opted not to try to apply further corrections to the ORS
selection functions besides those for extinction and for volume I of CGCG.

In deriving the selection functions, we correct diameters and
magnitudes for discreteness effects using the method described by
Strauss, Yahil, \& Davis (1991).  For the ESO-m sample, we take the
$1'$ diameter cut-off inherent to this sample into account when
deriving the selection function.  This is done by substituting $\min
[r_{max} (L_i), r_{max} (D_i)]$ for $r_{max} (L_i)$ in Equation (12)
for the likelihood function.  In all fits to \self, we assume that
distances are proportional to redshifts in the Local Group
rest frame, following Yahil, Tammann, \& Sandage
(1977) to convert from heliocentric velocities.  Only galaxies with $r
\leq 8000 \kms$ are included in each
fit, as the samples become unacceptably sparse at higher redshifts
(cf., Figure~4 of Paper 1).
Galaxies around rich clusters are placed at each
cluster's center, following Table~2 of YSDH.

Each galaxy is weighted by the inverse of the product of the selection
function and the mean density in each subsample, $n_1$. When combining
samples, we multiply by a further weight given by the inverse of the
relative density of \iras\ galaxies from the 1.2 Jy redshift survey
(Fisher \etal 1995) within the solid angle of each sample and within a
redshift of 8000 \kms. These relative densities are 1.025 for the UGC sample,
1.233 for the ESGC sample, 0.760 for the ESO sample limited in
diameter, and 0.785 for the ESO sample limited in magnitude (which
covers a smaller solid angle; see Paper 1).

The derived \self\ parameters for the various samples are given in
Table~1. They are labeled by the parent catalogue from which they were
drawn, with a suffix -d or -m for diameter and magnitude-limited
samples, respectively. We also list the number of galaxies used in the
maximum-likelihood fit, the derived extinction correction parameter
$\gamma$ and the global mean density for each sample. Formal fit
errors for the \self\ parameters typically vary from 10\% to
30\%. Errors for $\gamma$ are larger, especially for the
diameter-limited subsamples, where they can be larger than 100\%! This
is at least partially due to the fact that the approximation that
$\gamma_d$ is the same for all galaxies in the sample is not very
good (cf., Figure~1). Another factor may be systematic and random
errors in the diameters, which may be of larger amplitude than the
extinction effect. Of
course, there is also substantial covariance between the values of the
selection function parameters and $\gamma$. The ESGC-d and UGC-d subsamples
have $\gamma < 0$
(extinction increases the apparent size of the galaxy), which is
unphysical.

The errors in $\gamma_m$ for the ESO and UGC samples are 0.28 and
0.35, respectively, consistent with the error bars in the upper panel
of Figure~3.  The
$\gamma$ values for ESO-m and UGC-m are also consistent with unity, as
expected. But how much improvement does the incorporation of
extinction into the fit to
\self\ cause? The quantity $\Delta \log \Lambda$, the logarithm of the
ratio of the likelihood for the best value of $\gamma$ to that with
$\gamma$ set to zero, is shown in the last column of Table~1.  The
extinction corrections give a large improvement to the likelihood for
ESO-m. For UGC-m, the extinction corrections may be masked by the
larger errors in the Zwicky magnitudes.  For the diameter-limited
samples, the improvement is again small.

In Figure~8 we show the
quantity $n_1$\self, the number density of galaxies in each subsample
as a function of distance in the absence of inhomogeneities, for the
five ORS subsamples listed in the first half of Table~1.
Panel {\it a} refers to the diameter-limited subsamples, while panel
{\it b} shows the magnitude-limited subsamples. The effect of
extinction is quantified by showing, in the lower two panels, the
ratio of $n_1 \phi$ calculated without and with extinction taken into
account (the
parameters in the former case are not given in the tables).
For the magnitude-limited subsamples, the $\phi_{obs}$ are steeper than
the extinction corrected \self, in accordance with Figure~2b.  For the
diameter-limited subsamples, however, only ESO-d shows this
behavior. Given the larger uncertainties associated with $\gamma_d$,
we decided to fix $\gamma_d = 0.6$, a value appropriate for ESO and
UGC galaxies (Fig.~1).  The
derived \self\ parameters for these fits are listed in the second half
of Table~1.

Luminosity and diameter functions are plotted in Figure~9. The units
of the luminosity function are galaxies$\,$Mpc$^{-3}\,$mag$^{-1}$,
while for the diameter function they are
galaxies$\,$Mpc$^{-3}\,$(5 log D/kpc)$^{-1}$. Panel {\it a}
corresponds to the diameter-limited samples and panel {\it b} to the
magnitude-limited ones.  As mentioned earlier, these functions are
likely to be affected by Malmquist bias. Correcting for it would work
in the opposite sense as extinction corrections, making the luminosity
functions steeper, but unfortunately we do not have a sufficiently
accurate model of the magnitude and diameter error distribution
to carry out this correction. This makes it difficult to compare the
results presented here to previously published luminosity functions in
the literature.

The ESO, ESGC and UGC diameter functions are
systematically displaced from one another. This is due to two effects:
the fact that the ESO volume is underdense relative to the UGC (ESGC) volume
by 30\% (60\%), as given by the \iras\ redshift survey, and the fact that the
ESO diameters are systematically larger (smaller) than the UGC (ESGC)
diameters by some $25\%$ ($10\%$) (Paper 1).
If we correct for these two effects, all three
diameter functions show a much improved agreement.
However, the best agreement between ESO and UGC is obtained if we assume that
$D_{ESO} = 1.20 D_{UGC}$, slightly smaller a correction than in Paper 1.
The corrected diameter functions are shown in panel {\it c} for all three
subsamples. Also shown (solid line) is the diameter function of Hudson \&
Lynden-Bell (1991) derived from the CfA1 sample (Huchra \etal 1983) using
UGC diameters (where we put in the same correction factor of 1.2
used for UGC-d).
Notice that the revised scaling factor between ESO and UGC diameters
is in closer agreement with
Lynden-Bell, Lahav, \& Burstein (1989). These latter also
matched the UGC and ESO diameter functions and
found a diameter scaling ratio of $1.17 \pm 0.07$, but they did not
have as complete redshift information, and they did not correct for
the underdensity of the ESO volume. As for the ESGC, its corrected diameter
function is a bit steeper than the other two, which could be
caused by residual non-linearities in the diameters in this sample. Finally,
as the Hudson \& Lynden-Bell sample overlaps substantially with UGC-d, we are
not surprised to find good agreement.

The agreement between the ESO and UGC luminosity functions is
very good at bright magnitudes.
UGC is slightly steeper on the most luminous end. However, these plots
do not correct for the underdensity of ESO relative to the UGC; with
this correction made, in panel {\it d}, one can match the two
luminosity functions above the knee in the luminosity functions if one
assumes that the ESO magnitudes are systematically brighter by 0.2 mag
than those of the UGC. However, the agreement at the low-luminosity
remains poor. Again, this may be a sign of residual
non-linearities in the magnitudes in one of the two samples, or
perhaps differences in the corrections needed for Malmquist bias.
In panel {\it d}, we also show the luminosity function derived by Loveday
\etal (1992) from a sample with essentially no overlap with ours.
The agreement with UGC is very good over the entire range
of magnitudes, but we again had to apply an empirical shift of 0.2 mag to the
Loveday \etal data.

We now discuss the quality of our best fit solutions for the 5 ORS
subsamples considered. As discussed by YSDH, the fit quality can be
assessed by comparing the observed distribution of luminosities to
that predicted from the derived \lumf; for each galaxy at a given
distance $r$, the predicted distribution function of luminosity is
given by Equation~(12). Moreover, if \lumf\ is in fact
universal (which is an assumption inherent to the maximum-likelihood
technique used) and the fitted solution is reasonable, the predicted
distribution of luminosities should match that observed for any
subset of the sample without explicit selection on luminosity.

The histograms in Figure~10 are the observed luminosity and diameter
distributions for the 5 ORS subsamples
considered in this paper. The smooth curves
are the expected distributions, given by Equation~(12) and the
observed distribution of distances.

We quantify the difference between the observed and predicted
distributions by computing the statistic:
$$\chi^{2}/{\rm dof} = {{\sum_{bins} \bigl(N_{obs} - N_{pred}\bigr)^2 /
\sigma_{obs}^2} \over {{\rm dof}}}, \eqno (17)$$
where $N_{obs}$ and $N_{pred}$ are the observed and predicted number of
galaxies at each bin in absolute magnitude (or diameter), respectively,
{dof} is the
number of degrees of freedom available in the comparison, and
$\sigma_{obs} = N_{obs}^{1/2}$ is the Poisson noise in the observed counts.
As in YSDH, only bins with more than 5 objects were used in
the $\chi^2$ determination.
This quantity is shown in each panel of Figure~10, and is repeated in
Table~2. The ratios are close to unity, indicating a good quality fit.

In the remaining lines of Table~2 we show the $\chi^2 / {\rm dof}$
for different subsets of the data, defined according to
distance, density, extinction amplitude and morphological
type. The values of $\chi^2/{\rm dof}$ are in general consistent with
unity. There are some exceptions, however. The large
$\chi^2/{\rm dof}$ for the lowest \ab\ range in ESO-m, for instance,
could be indicative of problems in the BH extinction maps in the
Southern hemisphere. However, we would then expect a similarly poor
fit in ESO-d, which is not the case. The $\chi^2/{\rm dof}$ in the
last distance bin of UGC-m could be due to Malmquist bias.

There are also several large $\chi^2/{\rm dof}$ values for some ranges
in density which might indicate a dependence of
\lumf\ on environment density.
In fact, variations in $\Phi (L)$ have been found
as a function of local density, usually quantified by a change in the
correlation function of galaxies as a function of luminosity
(Maurogordato \& Lachi\`eze-Rey 1987; Hamilton 1988; Davis \etal 1988;
Bouchet \etal 1993; Park \etal 1994).  This issue is, however, still
subject to debate (Alimi, Valls-Gabaud, \& Blanchard 1988,
Valls-Gabaud, Alimi, \& Blanchard 1989).
Finally, several morphological subsamples show large values of
$\chi^{2}/{\rm dof}$, indicating that the luminosity function may be a
function of morphology. Again, dependence of \lumf\ on morphology has
been observed by several authors (e.g., Binggeli \etal 1988, Ferguson
\& Sandage 1991, Santiago \& Strauss 1992).
Notice, however, that no systematic trends
in the quality of the fits are found as a function of either extinction,
density, velocity or Hubble type, for both  diameter and magnitude-limited
subsamples.
Thus, we may conclude that the fits obtained to each subsample are
reasonable; moreover, our data are consistent with our assumption that
the luminosity and diameter functions are universal.

Extinction corrections should also depend on morphology
via differences in profile shapes as shown in \S 2.2 (cf., Cameron 1990).
In order to assess if this causes systematic errors in the selection
function, we derived
separate \self\ for different subsets defined according to the de Vaucouleurs
numerical T types. The selection function parameters, number of galaxies,
$\gamma$ and global mean densities for several morphological classes
are shown in Table~3. We restrict ourselves to ESO-m and UGC-m,
since we earlier found that extinction corrections for the diameter-limited
samples are very uncertain.

The fits are somewhat more uncertain than those
of Table~1 due to the smaller number of galaxies in each subsample.
We again use the $\chi^2/\dof$ statistics, listed in the last column of
Table~3, to quantify the difference between observed and expected
distributions. All, with the exception of the spirals in UGC-m, are close
to unity. Statistical noise is particularly important for
the Irregulars/Dwarfs subset ($7 \leq T \leq 11$), where in spite of the
good fit, the formal errors in the parameters are substantial.
Clear differences in the \lumf\ parameters are seen as a function
of type. In particular, $\gamma$ is systematically larger for late-type
galaxies than for early-types.
Notice, however, that the errors in $\gamma$ are large enough
to render most values of this parameter consistent with 1.
Furthermore, the existence of correlations among the parameters used in the
maximum-likelihood fit is also clear from Table~3. In particular,
$\beta$ and $r_*$ are strongly correlated. Figure~11 illustrates this
with likelihood contours for the entire ESO-m subsample
projected on the $\beta - r_*$ plane (panel {\it a}) and
the $\gamma - r_*$ plane (panel {\it b}). The two contours delimit the 68\% and
90\% confidence level regions in the parameter values. The correlation
between $\beta$ and $r_*$ stands out clearly. $\gamma$ is also mildly
correlated with $r_*$.

Finally, we use our best fit solutions to the luminosity function in
ESO-m, UGC-m and ESGC in order to estimate spatial galaxy densities.
In Paper 1 we showed density contours on constant redshift shells; in
Figure~12 we show planar slices parallel to the principal planes in
Supergalactic coordinates. Similar plots of the galaxy distribution
have been shown by Saunders \etal (1991), Strauss \etal (1992a),
Fisher \etal (1995) and Strauss \& Willick (1995) for \iras-selected
galaxies, and Hudson (1993) for optically-selected galaxies.  Gaussian
smoothing was used, with a width given by the mean interparticle
spacing of the UGC-m sample at each distance. The zone of avoidance
($|b| < 20^\circ$) is indicated by shading in each panel. The
calculations were done to a radius of 8000 \kms, indicated by the
dashed circle in each panel.

The Supergalactic Plane (SGZ${}= 0$) is the central panel in the plot.
Some of the more dramatic features are labeled here: the Virgo
cluster (V) is the overdensity near the origin. The Hydra-Centaurus
(H-C) and Pavo-Indus-Telescopium (P-I-T) superclusters lie just above
and below the zone of avoidance at SGX${}=-3000 \kms$. The \iras\
galaxy distribution, which continues to lower Galactic latitudes,
shows these two structures to be contiguous, together making up the
Great Attractor. At lower SGY is the Sculptor Void (Sc). The Coma
cluster (Co) lies at SGY${}=6500 \kms$, SGX${}=0$.  The Pisces-Perseus
(P-P) supercluster lies largely in the zone of avoidance, although
much of it is apparent at SGX${}=4500\kms$, SGY${}=-3000 \kms$; it is
contiguous with the Cetus supercluster at lower SGY.

Slices 3000 \kms\ above and below the Supergalactic plane are shown
in the two panels to the right and left of the central panel
respectively. The region above the Supergalactic plane is dominated by
a large void, and the high SGZ extension of the Great Attractor. The
most prominent structure in the slice at SGZ${}=-3000\kms$ is a part
of the Great Wall (Geller \& Huchra 1989) at SGX=${}-2000 \kms$,
SGY=${}6000 \kms$.

Slices at constant SGX are found in the upper three panels.
SGX${}=0$ (center) cuts through the Virgo and Coma clusters, while the
right and left panels cut through the Pisces-Perseus and
Hydra-Centaurus superclusters, respectively. Finally, the lower three
panels show slices at constant SGY. SGY${}=0$ lies completely in the
Galactic plane, and the SGY${}=-3000 \kms$ panel cuts through the
Southern part of the Pisces-Perseus supercluster, and the Sculptor
Void.

\bigskip
\centerline {\bf 4. Conclusions}
\medskip\nobreak
We have used Monte-Carlo simulations to assess the amplitude of
biases caused by different types of systematic and random errors
on the derived selection function and density field of a galaxy sample.
Our results may be summarized as follows:
\medskip
\item{1-} Galactic extinction causes a non-uniform selection effect
across the sky. The selection function derived in the usual way
(as a function of radial distance only) is incapable of accounting for
this effect. It at best reflects the mean loss
of galaxies averaged over all directions. The same conclusion applies
to any other systematic position-dependent bias present in the
magnitudes or diameters used in selecting a sample.
\medskip
\item{2-} An unbiased selection function can be recovered from
a redshift sample in the case of extinction, by taking this effect
explicitly into account.
However, other (unknown) systematic
as well as random errors on
diameters and magnitudes interfere with extinction corrections.
\medskip
\item{3-} The effect of extinction
on diameters is both smaller and more variable than that on magnitudes.
These two factors make it harder to
incorporate extinction corrections into the selection function derivation
for diameter-limited samples than for their magnitude-limited counterparts.
\medskip
\item{5-} The density field is not strongly affected by
systematic effects whose amplitudes do not vary across the sky, since
these can still be properly incorporated into a
radial selection function.
Furthermore, shell densities averaged over solid angle are not biased even in
the presence of
errors as a function of position on the sky, since they depend on radial
distance only and, as such, involve
an averaging of densities taken over the entire sample solid angle.
\medskip
\item{6-} Random errors also corrupt the derived selection function.
As in the case of systematic errors, however,
they significantly affect the density field
only if their typical amplitude varies across the sky.
\medskip
\item{7-} Luminosity and diameter functions are systematically biased
by errors in the magnitudes and diameters, respectively. Unbiased
estimates of \lumf\ and \diamf\ can be recovered only if these errors
are known and corrected for when deriving the selection function.
This applies to both uniform and non-uniform errors, either systematic
or random.
\medskip
Using the maximum-likelihood technique described in \S 2.2, extinction
corrected selection and luminosity
functions were derived for the ORS magnitude-limited
subsamples. These may, however, be biased by
systematic and random errors. For the diameter-limited subsample, the
effect of extinction was, as expected, harder to quantify,
leading to unacceptably
large uncertainties in the extinction corrections.
We thus chose to apply a fixed ``typical'' extinction correction for the
diameter-limited subsamples.

We derived separate selection functions for the three different catalogues from
which ORS was drawn. This was done in order to circumvent biases
caused by non-uniform sample selection among the three parts of the
sky. For comparison, we also derived radial selection functions that
completely ignore extinction. The amplitude of the extinction bias in
ESO-m and UGC-m is similar to that expected from the Monte-Carlo
simulations.  The derived selection functions were tested by comparing the
observed distribution of luminosities in each sample to that predicted
{}from the fit itself. This comparison was done for several subsets of
each ORS subsample. The agreement between observed and expected
distributions is good, implying consistency with the
implicit assumption that \lumf\ is universal. The luminosity and
diameter functions are consistent with recent determinations
in the literature.

Type-dependent selection functions were derived from ESO-m and
UGC-m subsamples. We find a trend of increasing $\gamma$ for later
morphological types. However, this result is tentative, given the
covariance between the selection function parameters and $\gamma$.

Finally, the ORS density field was plotted in planes parallel to the
Supergalactic axes. As discussed in the previous sections, this
density field is fairly unbiased by random and systematic errors,
apart from position-dependent systematic trends or strong and variable
scatter in the magnitudes or diameters.  In the next paper of this
series (Strauss \etal 1995) we will make quantitative comparisons of
the density field as traced by different morphological subsets of the
ORS, and well as comparisons of the density field to that of
infrared-selected galaxies.

{\bf Acknowledgments.} We thank Karl Fisher for his help in generating
the $N$-body models. MAS is supported at the IAS under a grant from
the W.M. Keck Foundation. MD acknowledges support of NSF grant
AST-9221540.  This research has made use of the NASA/IPAC
Extragalactic Database (NED) which is operated by the Jet Propulsion
Laboratory, Caltech, under contract with the National Aeronautics and
Space Administration.  BXS acknowledges a doctoral fellowship from the
{\it Conselho Nacional de Desenvolvimento Cient\'\i fico e
Tecnol\'ogico} (CNPq), and the generous hospitality of the Caltech
Astronomy department. BXS and OL thank the Institute for Advanced
Study for its invitations to visit. This work is partially based on
the PhD. thesis of BXS from Observatorio Nacional, Rio de Janeiro,
Brazil.

\vfill\eject

\centerline{{\bf Table~1}.
ORS Selection Function Parameters}
\vskip 0.2 true cm
\hrule
\vskip 0.2 true cm
\tabskip=1em plus2em minus.5em
\halign to\hsize
{#\hfil&&\hfil#&\hfil#&\hfil#&\hfil#&\hfil#&\hfil#&\hfil#&\hfil#\cr
Sample & $N_g$ & $\alpha$ & $\beta$ & $r_*^a$ &
$\gamma$ & $n_1^b$ & $\Delta \log \Lambda $ \cr
\noalign{\smallskip\hrule\smallskip}
ESO-m &  2170 & 0.40 & 6.25 & 11100 & 1.00 & 10.88 & 23.2 \cr
ESO-d &  1614 & 0.39 & 3.19 & 6375 & 0.60 & 6.87 & 1.9 \cr
UGC-m &  2965 & 0.36 & 9.38 & 14060 & 0.56 & 8.08 & 0.9 \cr
UGC-d &  1848 & 0.33 & 2.83 & 4075 & $-$0.32 & 8.60 & 0.5 \cr
ESGC-d & 1200 & 0.47 & 5.26 & 9500 & $-$0.34 & 18.24 & 0.8 \cr
\noalign{\smallskip\hrule\smallskip}
ESO-d &  1614 & 0.39 & 3.19 & 6375 & 0.60 & 6.88 & 1.8 \cr
UGC-d &  1848 & 0.33 & 2.76 & 4115 & 0.60 & 9.63 & $-$2.1 \cr
ESGC-d & 1200 & 0.50 & 5.44 & 10565 & 0.60 & 20.20 & $-$3.2 \cr
}
\smallskip\hrule
\smallskip
\noindent Note: all samples limited to 8000 \kms.

\noindent $^a$ \kms\

\noindent $^b$ $10^{-2}$ $h^{-3}$ Mpc$^{-3}$. No global correction
of density from \iras\ applied.

\vfill\eject

 \centerline{{\bf Table~2}.
$\chi^{2} / {\rm dof}$ for the comparison between predicted}
\smallskip
\centerline{and observed luminosity (or diameter)
distributions of ORS subsamples}
\vskip 0.2 true cm
\hrule
\vskip 0.2 true cm
\tabskip=1em plus2em minus.5em
\halign to\hsize
{#\hfil&\hfil#&\hfil#&\hfil#&\hfil#&\hfil#\cr
Subsample & ESO-m & ESO-d & UGC-m & UGC-d & ESGC-d \cr
\noalign{\smallskip\hrule\smallskip}
All  & 33.6/25 & 37.0/25 & 27.0/27 & 27.2/28 & 22.5/22 \cr
\noalign{\medskip}
V $\leq$ 2000  & 24.1/21 & 41.1/21 & 15.6/22 & 24.1/22 & 7.6/13 \cr
2000 $\leq$ V $\leq$ 4000  & 14.5/15 & 19.8/14 & 17.0/15 & 10.6/11 &
17.1/12 \cr
4000 $\leq$ V $\leq$ 6000  & 7.5/9 & 5.4/9 & 17.9/10 & 1.8/10 & 11.4/9 \cr
6000 $\leq$ V $\leq$ 8000  & 9.5/7 & 4.6/6 & 29.6/8 & 4.6/5 & 8.4/6 \cr
\noalign{\medskip}
$\delta \leq$ -0.5  & 20.9/23 & 45.8/24 & 21.6/26 & 30.5/27 & 15.6/20 \cr
- -0.5 $\leq \delta \leq$ 0.5 & 3.5/6 & 2.2/2 & 6.8/11 & 1.9/6 & 2.7/6 \cr
0.5 $\leq \delta \leq$ 2.0 & 7.5/12 & 9.2/8 & 19.5/14 & 9.8/12 & 14.7/7 \cr
$\delta \geq$ 2.0 & 15.2/15 & 11.0/15 & 20.8/22 & 21.0/21 & 20.7/9 \cr
Clusters & $-/-$ & $-/-$ & 10.0/15 & 20.3/11 & $-/-$ \cr
\noalign{\medskip}
$A_B \leq 0.05$  & 51.5/21 & 22.0/22 & 32.8/25 & 37.7/26 & 15.2/16 \cr
$0.05 \leq A_B \leq 0.10$  & 10.8/13 & 8.9/12 & 11.5/17 & 12.8/19 & 22.9/17 \cr
$0.10 \leq A_B \leq 0.25$  & 14.1/16 & 30.3/16 & 13.0/20 & 17.2/21 &
12.4/16 \cr
$0.25 \leq A_B \leq 1.00$  & 12.2/15 & 13.8/16 & 11.9/10 & 8.6/12 & 7.7/11 \cr
\noalign{\medskip}
$T \leq 0$  & 34.3/22 & 13.0/17 & 28.0/22 & 9.8/19 & 30.4/15 \cr
$1 \leq T \leq 6$  & 26.5/20 & 40.7/19 & 26.1/23 & 31.9/22 & 11.6/15 \cr
$7 \leq T \leq 11$  & 25.0/14 & 34.0/16 & 48.8/19 & 26.1/20 & 15.3/14 \cr}
\smallskip\hrule
\vfill\eject

\centerline{{\bf Table~3}.
ORS Selection Function Parameters for Different Morphological Types}
\vskip 0.2 true cm
\hrule
\vskip 0.2 true cm
\tabskip=1em plus2em minus.5em
\halign to\hsize
{#\hfil&\hfil#&\hfil#&\hfil#&\hfil#&\hfil#&\hfil#&\hfil#\cr
Sample & $N_g$ & $\alpha$ & $\beta$ & $r_*^a$ &
$\gamma$ & $n_1^b$ & $\chi^2/{\rm dof} $ \cr
\noalign{\smallskip\hrule\smallskip
\centerline {ESO-m}\medskip}
T $\leq$ $-$3 ~~~~~  &  381 & 0.38 & 7.49 & 15540 & 0.51 & 1.07 & 14.2/16 \cr
$-$2 $\leq$ T $\leq$ 0 &  477 & 0.35 & 8.96 & 11880 & 1.01 & 3.09 & 16.8/17 \cr
1 $\leq$ T $\leq$ 4  &  801 & 0.20 & 14.71 & 17220 & 1.57 & 2.29 & 22.8/20 \cr
5 $\leq$ T $\leq$ 6  &  322 & 0.12 & 3.47 & 6090 & 1.00 & 0.76 & 11.0/20 \cr
7 $\leq$ T $\leq$ 11 &  189 & $-$0.05 & 2.42 & 1860 & 3.12 & 4.96 & 12.6/15 \cr
\noalign{\bigskip
\centerline {UGC-m}
\medskip}
T $\leq$ $-$3 ~~~~~  &  401 & 0.29 & 40.07 & 35670 & 0.42 & 0.57 & 13.9/18 \cr
$-$2 $\leq$ T $\leq$ 0 &  590 & 0.26 & 8.81 & 11520 & $-$0.19 & 1.71 &
19.2/24 \cr
1 $\leq$ T $\leq$ 4  &  961 & 0.00 & 2.93 & 5190 & 1.00 & 1.16 & 40.9/25 \cr
5 $\leq$ T $\leq$ 6  &  680 & 0.33 & 4.85 & 8730 & 0.85 & 1.90 & 23.0/25 \cr
7 $\leq$ T $\leq$ 11 &  215 & 0.86 & 8.27 & 10050 & 0.92 & 9.45 & 15.2/21 \cr}
\smallskip\hrule
\smallskip
\noindent Note: all samples limited to 8000 \kms.

\noindent $^a$ \kms\

\noindent $^b$ $10^{-2}$ $h^{-3}$ Mpc$^{-3}$. No global correction
of density from \iras\ applied.
\vfill\eject

\centerline {\bf REFERENCES}
\def\pp{\parshape 2 0truecm 13.4truecm 2.0truecm 11.4truecm}
\def\apjref #1;#2;#3;#4 {\par\pp#1, #2, #3, #4 \par}
\def\apj{\apjref}
\medskip
\hyphenation{MNRAS}
%\sevenrm

\pp Alimi, J.M., Valls-Gabaud, D., \& Blanchard, A.
1988, A\&A, 206, L11

\pp Binggeli, B., Sandage, A., \& Tammann, G.A. 1988, ARA\&A, 26, 509

\par\pp Blumenthal, G. R., Faber, S. M., Primack, J. R., \& Rees,
M. J. 1984, Nature, 311, 517

\par\pp Bouchet, F. R., Strauss, M., Davis, M., Fisher, K. B.,
Yahil, A., \& Huchra, J. P. 1993, ApJ, 417, 36

\par\pp Burstein, D., \& Heiles, C. 1982, AJ, 87, 1165

\pp Cameron, L.M. 1990, A\&A, 233, 16

\pp Choloniewski, J. 1985, MNRAS, 214, 197

\pp Choloniewski, J. 1986, MNRAS, 223, 1

\par\pp Cole, S., Arag\'on-Salamanca, A., Frenk, C. S., Navarro,
J. F., \& Zepf, S. E. 1994, MNRAS, 271, 781

\par\pp Corwin, H.G, \& Skiff, B.A. 1995, Extension to the Southern
Galaxies Catalogue, in preparation

\par\pp da Costa, L.N. \etal 1994, ApJ, 424, L1

\apj Davis, M., Geller, M.J, \& Huchra, J. 1978;ApJ;221;1

\pp Davis, M., \& Huchra, J. 1982, ApJ, 254, 437

\pp Davis, M., Meiksin, A., Strauss, M.A., da Costa, N., \& Yahil, A.
1988, ApJ, 333, L9

\pp Dekel, A., \& Rees, M.J. 1987, Nature, 326, 455

\par\pp de Lapparent, V., Geller, M. J., \& Huchra, J. P. 1989, ApJ,
343, 1

\pp de Vaucouleurs, G., de Vaucouleurs, A., Corwin, H., Buta, R., Paturel,
G, \& Fouqu\'e, P. 1991, {\sl Third Reference Catalogue of Bright
Galaxies} (New York: Springer)

\pp Efstathiou, G., Ellis, R.S., \& Peterson, B.A. 1988, MNRAS,
232, 431

\pp Ferguson, H.C., \& Sandage, A. 1991, AJ, 101, 765

\par\pp Fisher, K. B., Huchra, J. P., Davis, M., Strauss, M. A.,
Yahil, A., \& Schlegel, D.  1995, ApJS, 100, 69

\par\pp Geller, M. J., \& Huchra, J. P. 1989, Science, 246, 897

\pp Hamilton, A.J.S. 1988, ApJ, 331, L59

\pp Huchra, J. 1976, AJ, 81, 952

\pp Huchra, J. P., Davis, M., Latham, D., \& Tonry, J. 1983, ApJS, 52, 89

\par\pp Hudson, M. 1993, MNRAS, 265, 43

\par\pp Hudson, M.J., \& Lynden-Bell, D. 1991, MNRAS, 252, 219

\pp Kirshner, R.P., Oemler, A., \& Schechter, P.L.,
 1979, AJ, 84, 951

\pp Kirshner, R.P., Oemler, A., Schechter, P.L., \& Shectman, S.
1983, AJ, 88, 1285

\par\pp Koo, D., \& Kron, R. 1992, ARA\&A, 30, 613

\par\pp Kron, G.E., \& Shane, C.D. 1976, A\&SS, 39, 401

\apjref Lahav, O., Rowan-Robinson, M., \& Lynden-Bell, D. 1988;MNRAS;234;677

\par\pp Lauberts, A. 1982, {The ESO-Uppsala Survey of the ESO(B) Atlas}
(M\"unchen: European Southern Observatory)

\par\pp Lauberts, A., \& Valentijn, E. 1989, {\it The Surface Photometry
Catalogue of the ESO-Uppsala Galaxies} (M\"unchen: European Southern
Observatory)

\apj Loveday, J., Peterson, B.A., Efstathiou, G., \& Maddox, S.J.
1992;ApJ;390;338

\par\pp Loveday, J., Maddox, S. J., Efstathiou, G., \& Peterson, P. A.
1995, ApJ, 442, 457

\apj Lynden-Bell, D., Lahav, O., \& Burstein, D. 1989;MNRAS;241;325

\pp Maia, M.A.G., \& da Costa, L.N. 1990, ApJ, 352, 457

\par\pp Marzke, R. O., Huchra, J. P., \& Geller, M. J. 1994, ApJ,
428, 43

\pp Maurogordato, S., \& Lachi\`eze-Rey, M. 1987, ApJ, 320, 13

\pp Nicoll, J.F., \& Segal, I.E. 1983, A\&A, 118, 180

\par\pp Nilson, P. 1973, {Uppsala General Catalogue of Galaxies}, {
Uppsala Astron.\ Obs.\ Ann.}, {6}

\par\pp Park, C., Vogeley, M. S., Geller, M. J., \& Huchra, J. P. 1994,
ApJ, 431, 569

\par\pp Paturel, G. 1977, A\&A, 56, 259

\apj Paturel, G., Fouqu\'e, P., Lauberts, A., Valentijn, E.A., Corwin, H.G.,
\& de Vaucouleurs, G. 1987;A\&A;184;86

\apj Paturel, G., Fouqu\'e, P., Buta, R., \& Garcia, A.M. 1991;A\&A;243;319

\pp Sandage, A., Tammann, G.A., \& Yahil, A. 1979, ApJ, 232, 352

\par\pp Santiago, B.X., \& Strauss, M.A. 1992, ApJ, 387, 9

\pp Santiago, B.X., Strauss, M.A., Lahav, O., Davis, M., Dressler, A.,
\& Huchra, J.P. 1995, ApJ, 446, 457 (Paper 1)

\par\pp Saunders, W., Frenk, C., Rowan-Robinson, M., Efstathiou, G.,
Lawrence, A., Kaiser, N., Ellis, R., Crawford, J., Xia, X. Y., \&
Parry, I. 1991, Nature, 349, 32

\apj Schechter, P.L. 1976;ApJ;203;297

\pp Sodr\'e, L., \& Lahav, O. 1993, MNRAS, 260, 285

\par\pp Strauss, M. A., Davis, M., Yahil, A., \& Huchra, J. P. 1990,
ApJ, 361, 49

\par\pp Strauss, M.A., Davis, M., Yahil, A., \& Huchra, J.P. 1992a, ApJ,
385, 421

\par\pp Strauss, M. A., Huchra, J. P., Davis, M., Yahil, A., Fisher,
K. B., \& Tonry, J. 1992b, ApJS, 83, 29

\par\pp Strauss, M. A., Yahil, A., Davis, M., Huchra, J. P., \& Fisher,
K. B. 1992c, ApJ, 397, 395

\par\pp Strauss, M.A., \& Willick, J. 1995, Physics Reports, in press

\pp Strauss, M.A., Yahil, A., \& Davis, M. 1991, PASP, 103, 1012

\pp Strauss, M.A., \etal 1995, in preparation (Paper 3)

\pp Turner, E.L. 1979, ApJ, 231, 645

\pp Valls-Gabaud, D., Alimi, J.M., \& Blanchard, A. 1989,
Nature, 341, 215

\par\pp White, S. D. M., \& Frenk, C. S. 1991, ApJ, 379, 52

\pp Yahil, A. 1988, in Large Scale Motions in the Universe: A
Vatican Study Week, edited by V. C. Rubin, \& G. V. Coyne, S. J.
(Princeton: Princeton University Press), p. 219

\apj Yahil, A., Tammann, G.A., \& Sandage, A. 1977;ApJ;217;903

\par\pp Yahil, A., Strauss, M.A., Davis, M., \& Huchra, J.P. 1991,
{ApJ}, 372, 380 (YSDH)

\par\pp Zwicky, F. \etal 1961--1968, {\it Catalog of Galaxies and Clusters of
Galaxies}, in six volumes (Pasadena: California Institute of
Technology) (CGCG)

\vfill\eject

\def\fig #1, #2, #3, #4 {\smallskip\leftskip2.5em \parindent=0pt
#4}
%The following macro will give inserted figures. Comment it out if you
%don't have psfig.
\input psfig
\def\fig #1, #2, #3, #4 {
\topinsert
\smallskip
\centerline{\psfig{figure=#1,height=#2 in,width=#3 in}}
\medskip
{\smallskip\leftskip2.5em \parindent=0pt
#4}
\endinsert}

%\centerline {\bf Figure Captions}

\fig 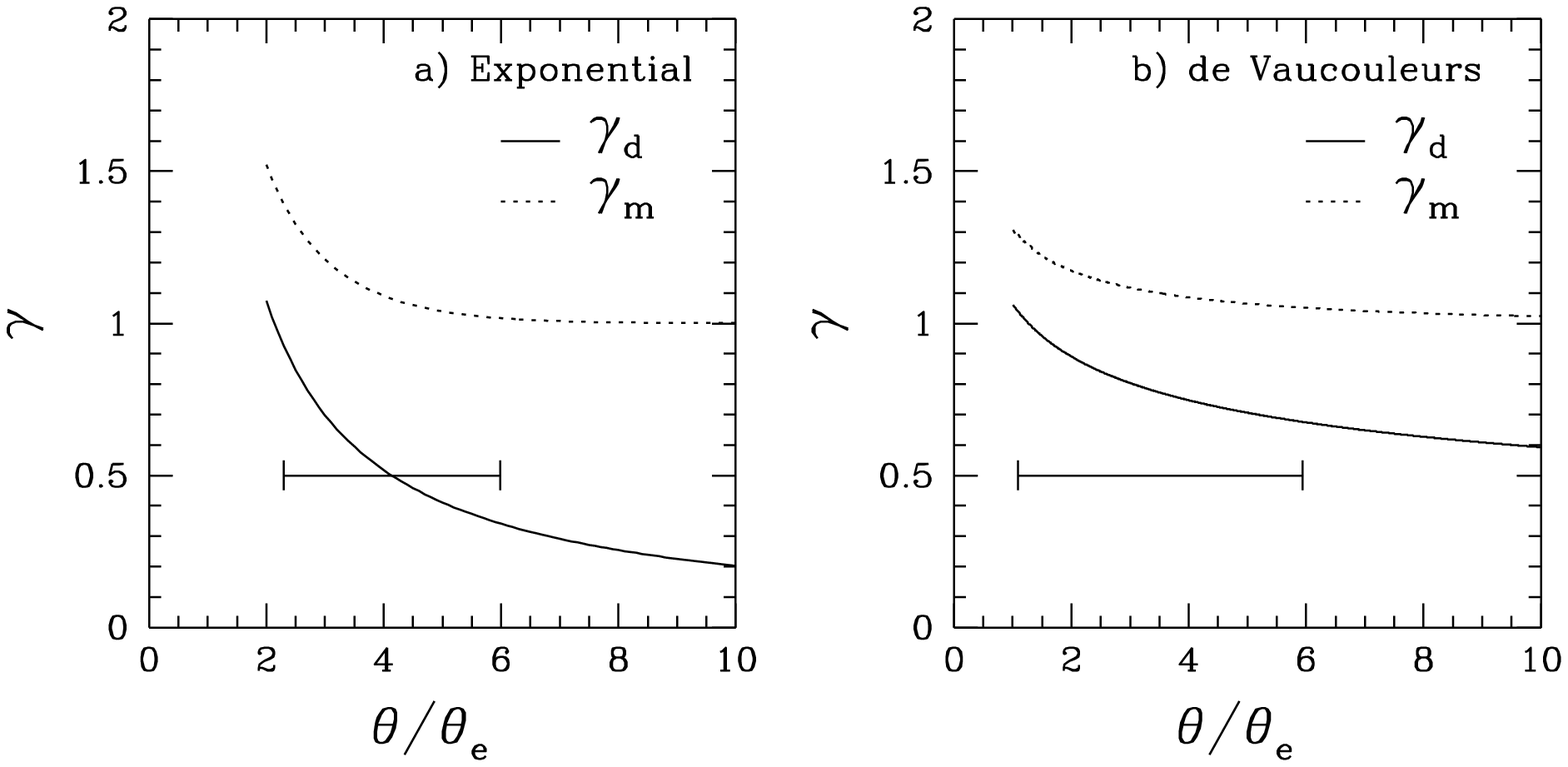, 5, 5, {1-  {\it a)} $\gamma$ values associated
with fluxes ($\gamma_m$, dotted line) and diameters
($\gamma_d$, solid line) for objects whose light profile
is an exponential and which are subject
to an extinction of $A_B = 0.3$ mag. $\gamma$ is shown
as a function of the number of $e$-folding
lengths $\theta/ \theta_e$.
The horizontal bars give the typical observed range of
$e$-folding lengths for ORS galaxies.
{\it b)} As in  {\it a}, for
a de Vaucouleurs profile. In this case, $\theta_e$ is the de
Vaucouleurs radius.}

\fig 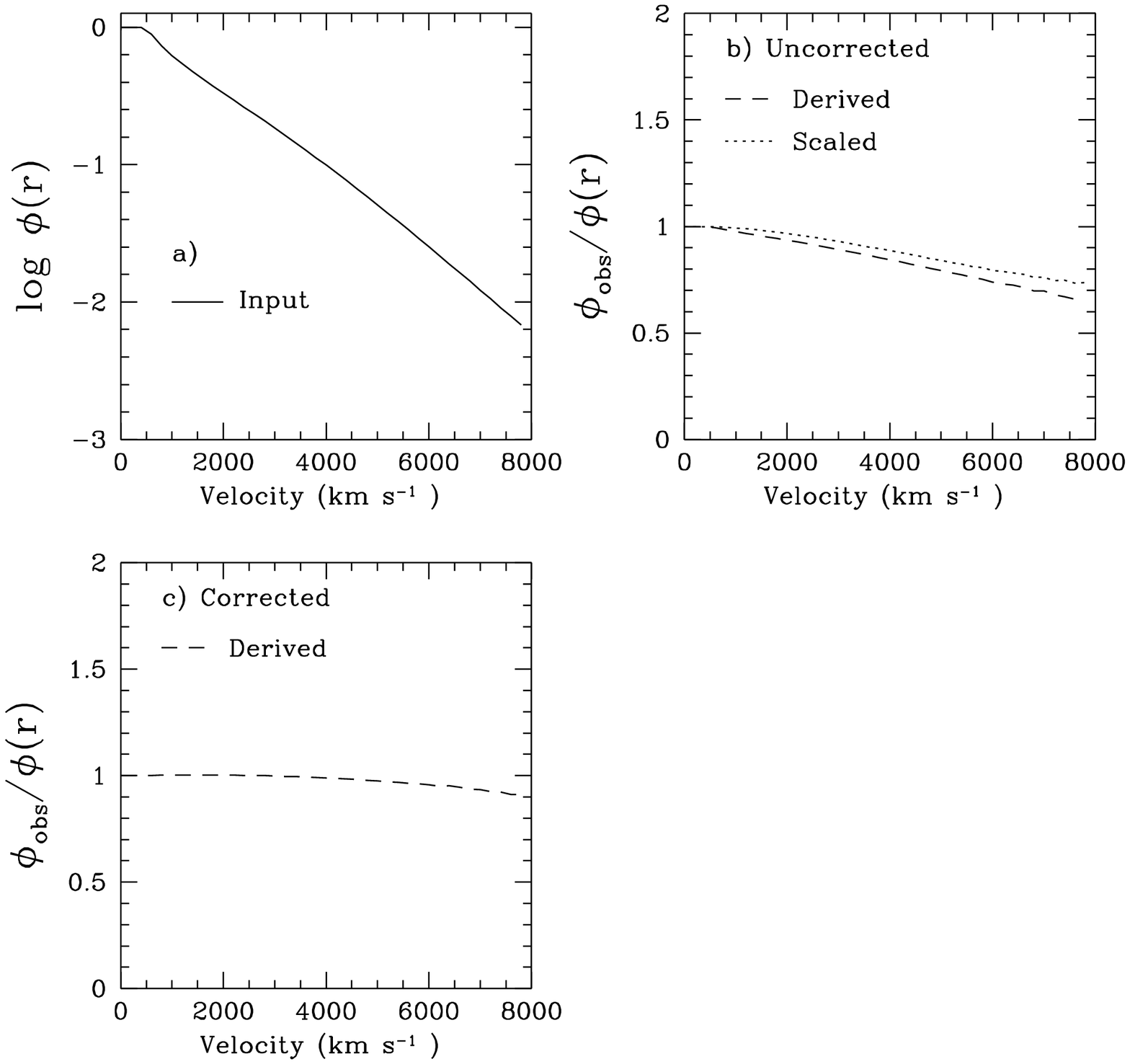, 5, 5, {2-  {\it a)} The input selection function
$\phi(r)$ used
for the Monte-Carlo simulations in this paper. Note the logarithmic abscissa.
{\it b}) The effect of
Galactic absorption on the selection function. The dashed line
represents the ratio of the derived to input selection function when
extinction is completely ignored, in a simulation with $\langle \gamma
\rangle = 1$ and extinctions drawn from Burstein \& Heiles (1982). The
dotted line is the selection function ratio, where the selection
function has been scaled using the mean $A_B$ value in Equation (11).
Note the linear scale.
{\it c)} The ratio of the selection function derived taking extinction
into account, to $\phi$.}

\fig 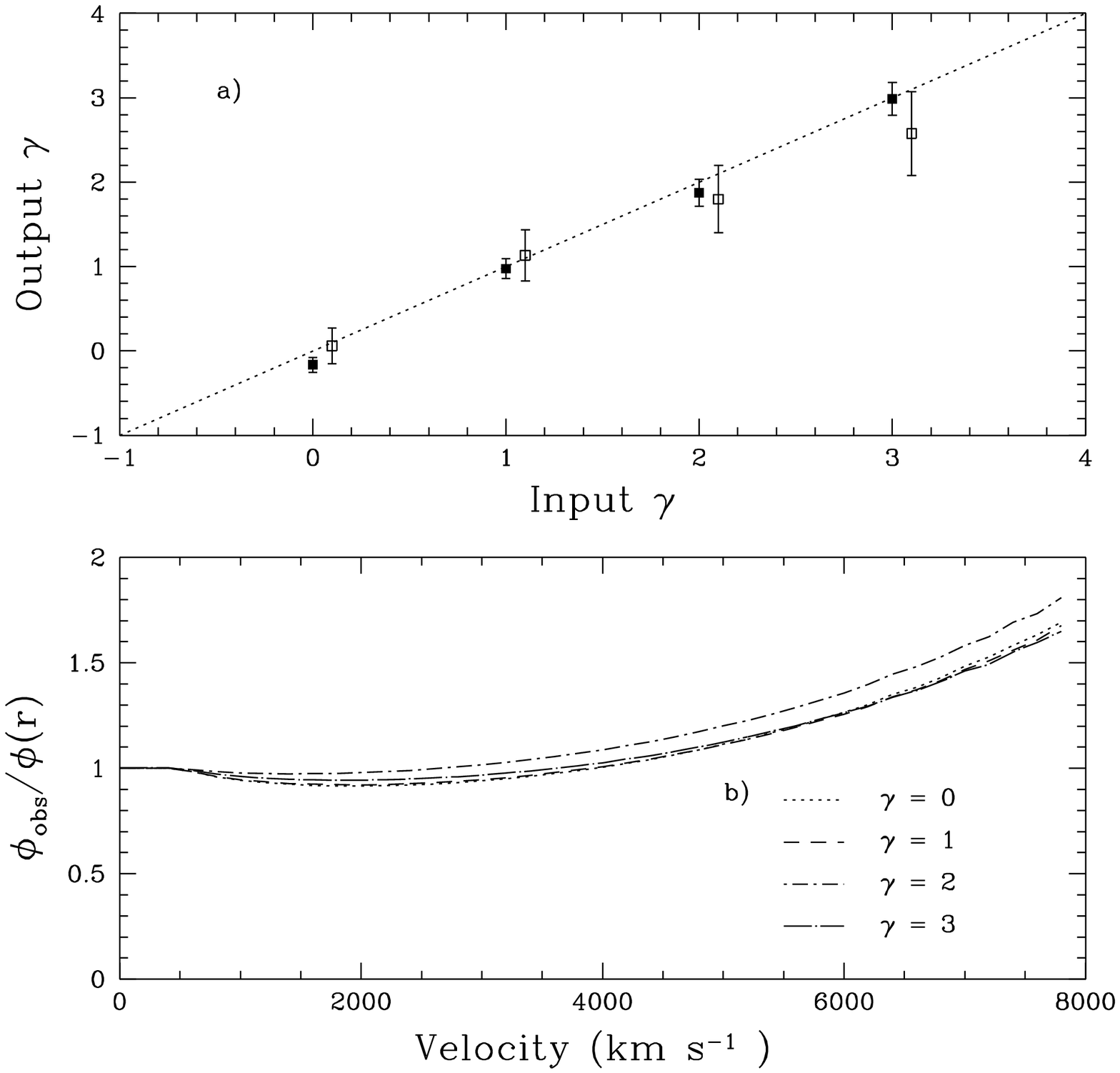, 5, 5, {3- {\it a}) Relation between the input
and derived values of the extinction
correction parameter $\gamma$ from Monte-Carlo simulations. The solid points
correspond to samples containing over 10000 galaxies.
The open symbols
refer to samples comparable in size to the ORS subsamples discussed
in \S 3 (about 2000 galaxies). Error bars are the
formal uncertainties from the maximum-likelihood
fit. {\it b}) Ratio of derived to input
selection function for the ORS-sized simulations whose
observed and input $\gamma$ values are given by the open points
in panel {\it a}.}

\fig 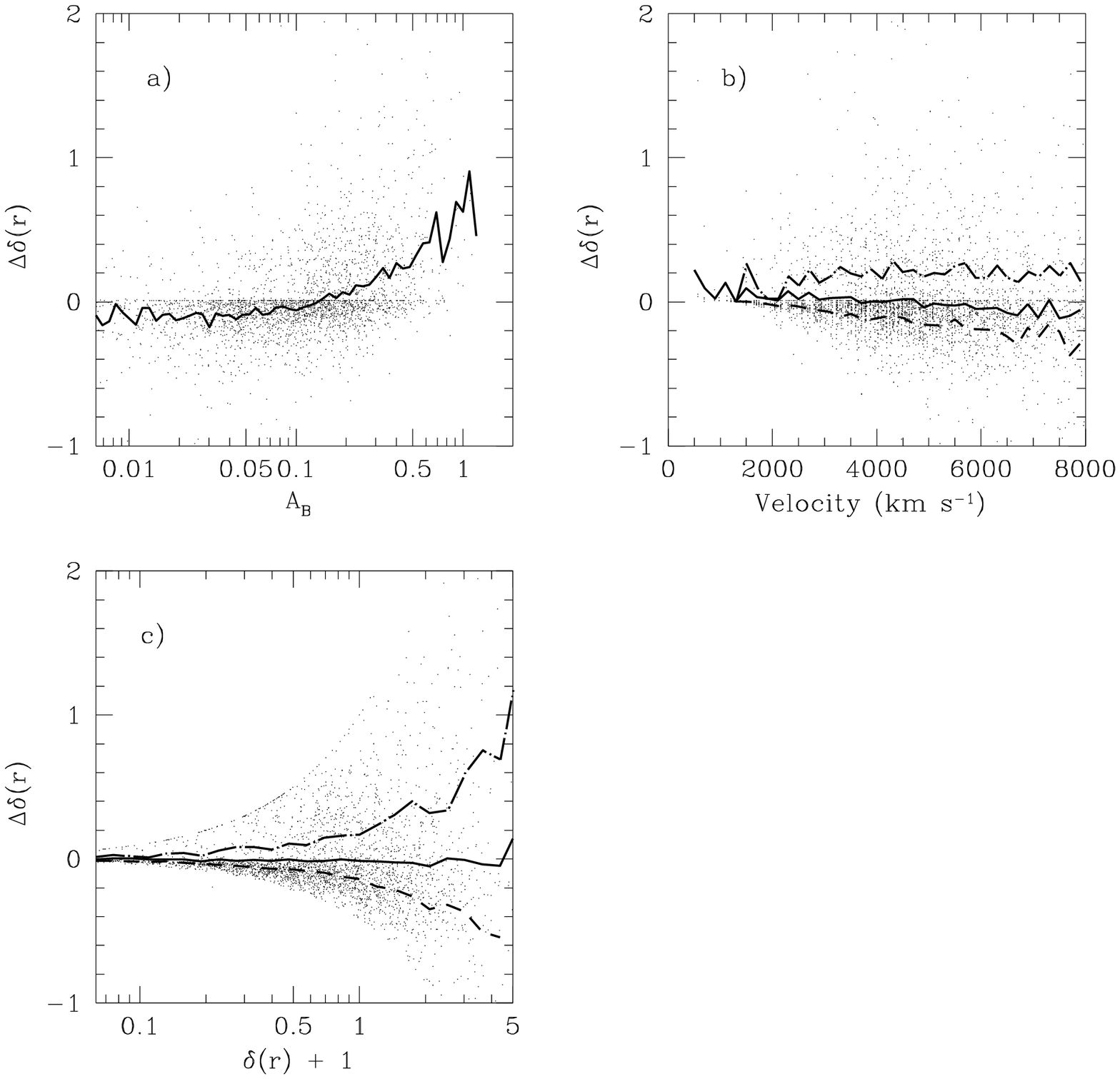, 5, 5, {4- The effect of Galactic absorption on the derived
density field.
The three panels give the difference between the input and
derived densities calculated on a grid of points for a Monte-Carlo
sample with input $\langle \gamma \rangle = 1$. The simulated
galaxies were weighted by the inverse of the derived selection function given
in Figure~2b. For clarity, only every second point is plotted.   {\it a)}
Density error as function of $A_B$.  The
solid line is the mean error in each bin of $A_B$.  {\it b)}
Density error as a function of
distance.  The solid line is the mean error in each bin of distance. The
dashed line (dot-dashed line) is the mean error over the
points with the 25\% lowest (highest) $A_B$ values.  {\it c)} As
in  {\it b}, now as a function of the input density. The upper
and lower envelopes in this case are due to the fact that both
$\delta$ and $\delta_{obs}$ have lower limits of $-1$.}

\fig 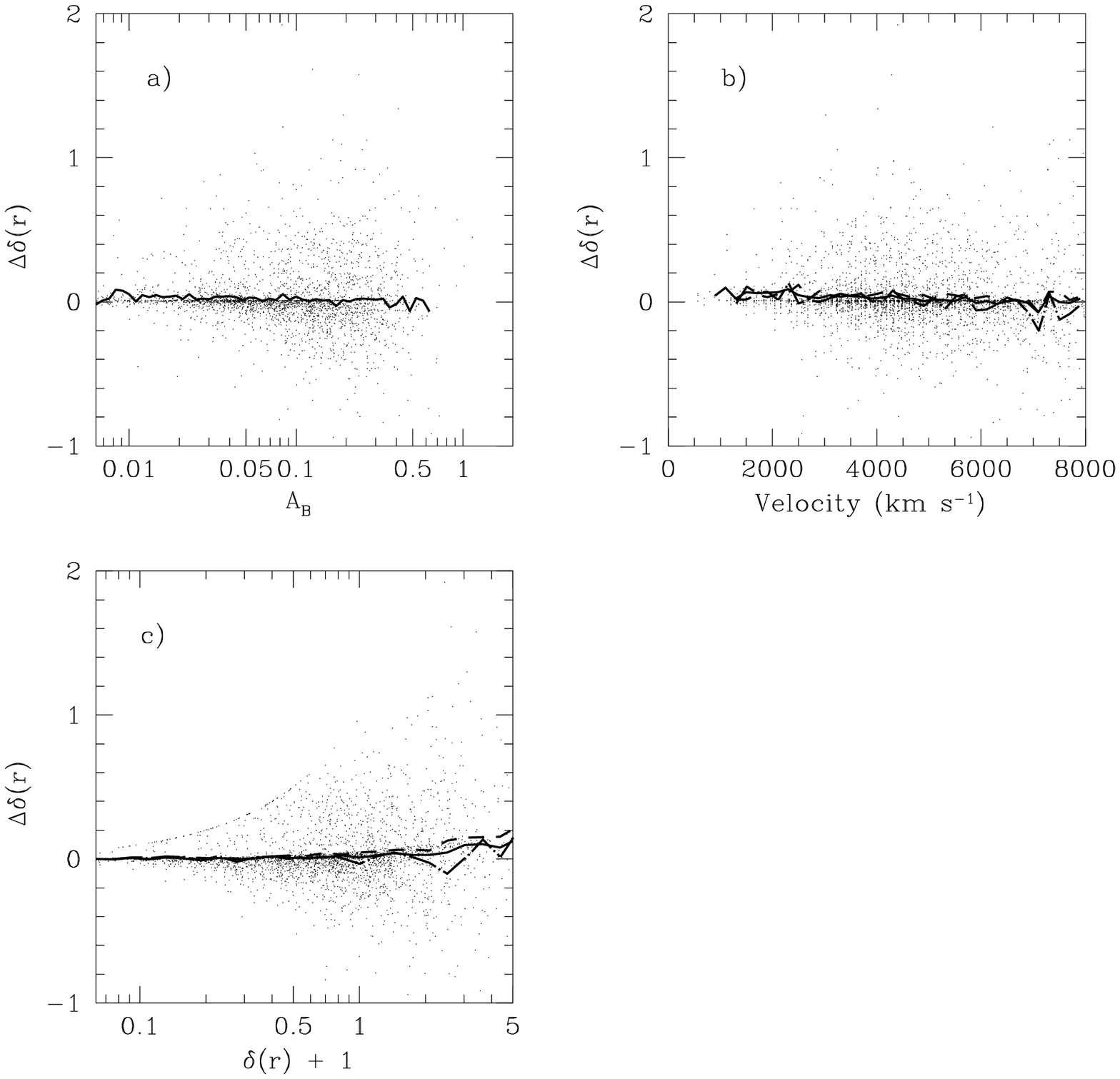, 5, 5, {5- As in Figure~4, with extinction properly corrected
for in the density calculation. }

\fig 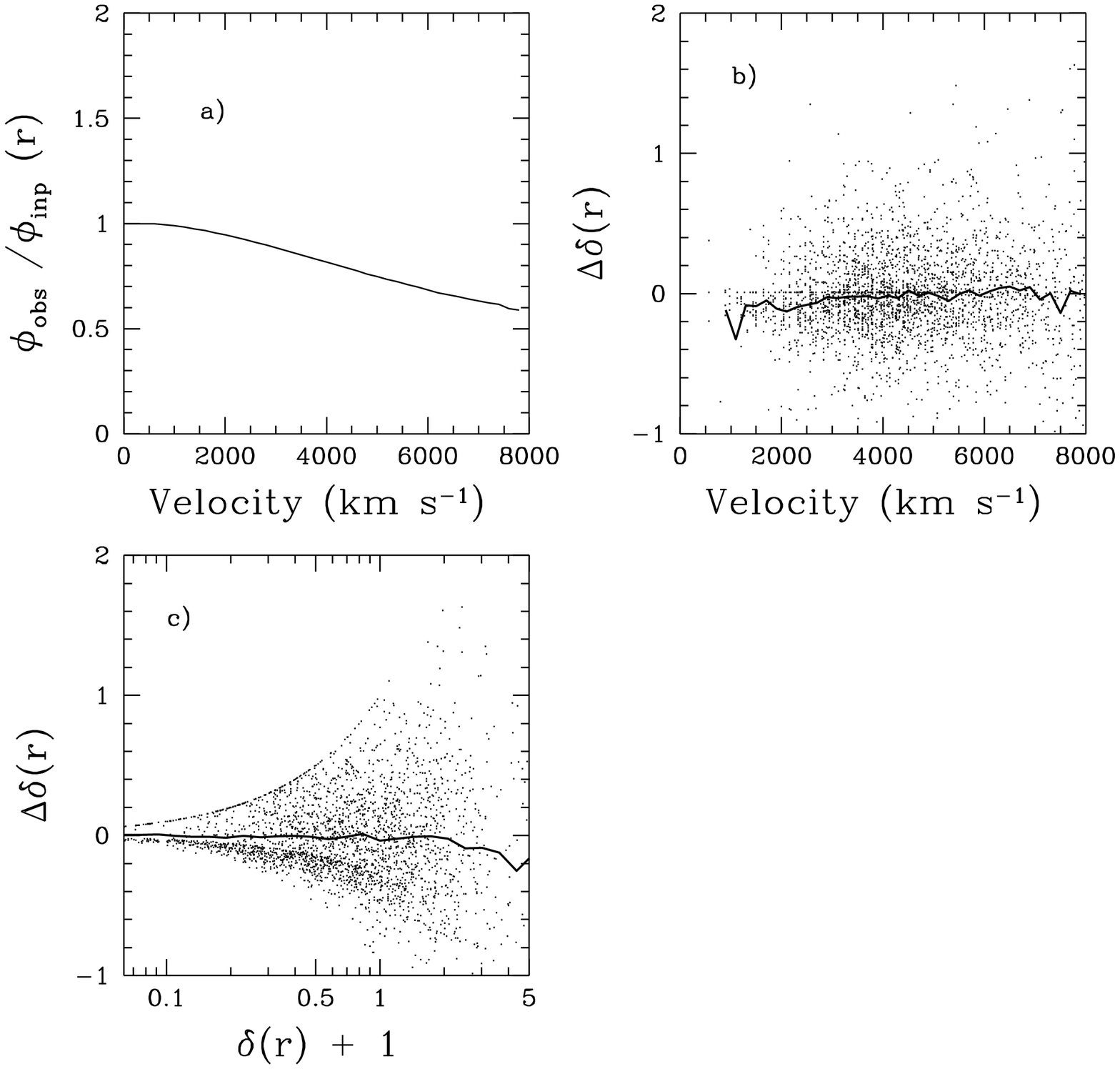, 5, 5, {6-  {\it a)} Ratio of derived to input selection
functions for a Monte-Carlo simulation whose magnitudes were systematically
biased according to Equation (13) in the text.  {\it b}) Difference
between input and derived densities as a function of velocity
distance. The solid line is the mean difference in distance bins.
 {\it c}) Errors in density plotted as a function of input density. }

\fig 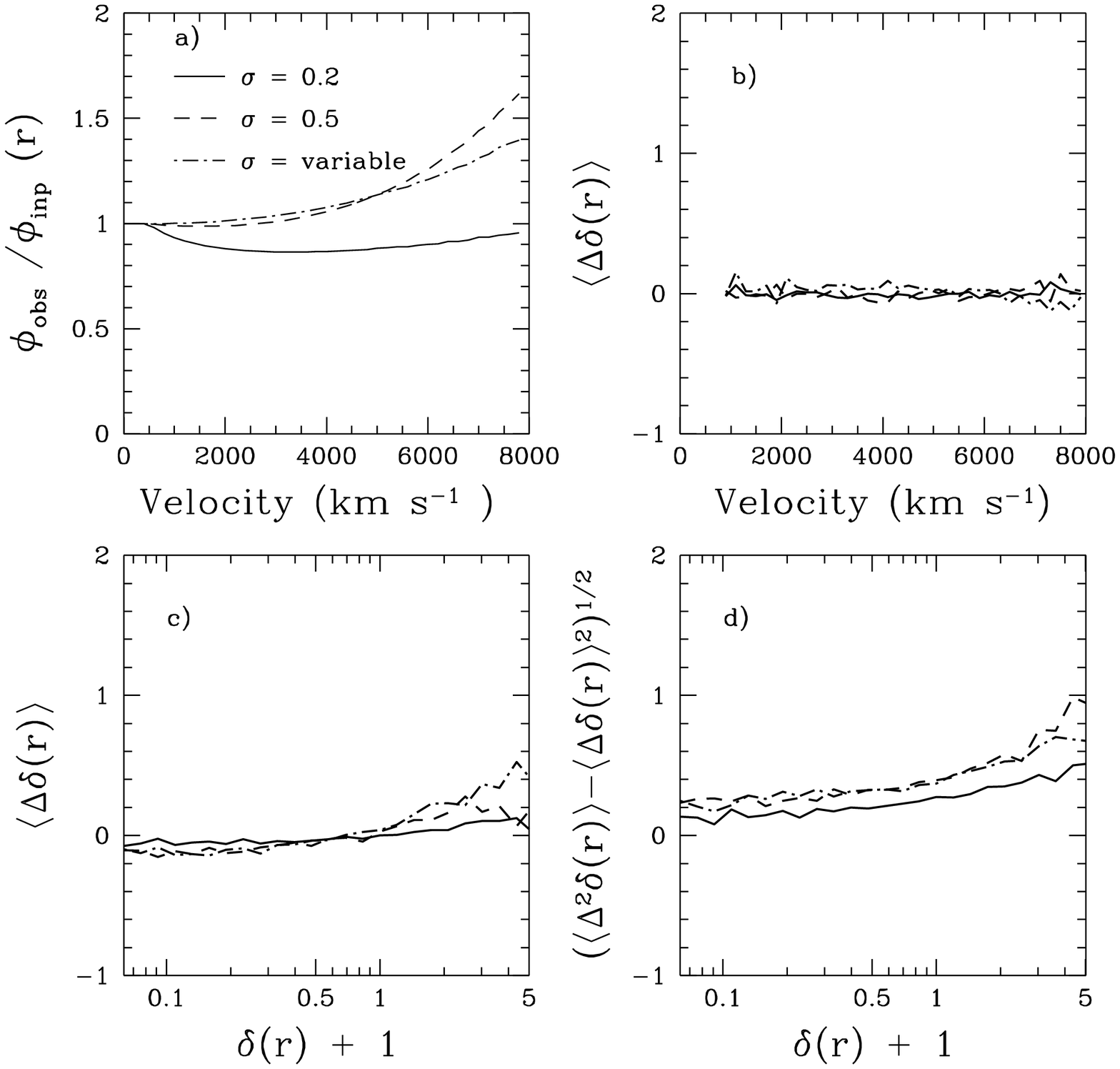, 5, 5, {7-  {\it a)} Ratio of derived to
input \self\ for three Monte-Carlo simulations
whose magnitudes were subjected to
random errors Gaussian distributed with standard deviations 0.2 mag (solid),
0.5 mag (dashed) and linearly increasing with magnitude (dot-dashed).
 {\it b)} Mean error in the density field as a function of
velocity for the three Monte-Carlo simulations whose selection functions
are given in {\it a}.  {\it c)} The mean error for the three
simulations as a function of the input density. {\it d} The rms error
for the three simulations as a function of the input density.}

\fig 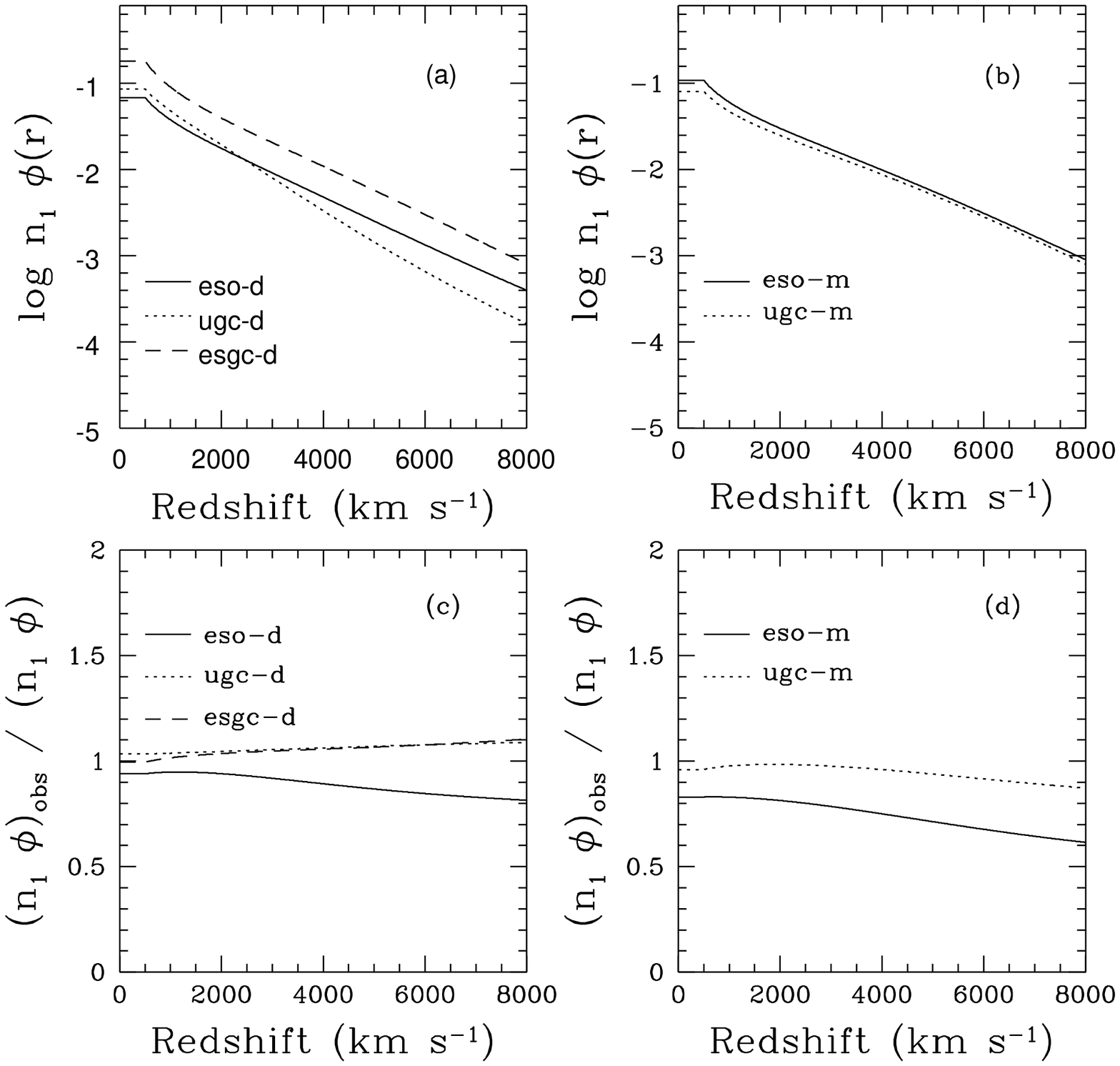, 5, 5, {8- {\it a)} Product of derived mean density and
selection functions for the diameter-limited subsamples: ESO-d (solid
line), UGC-d(dotted line) and ESGC-d (dashed line).
{\it b)} As in {\it a} for the magnitude-limited subsamples.
{\it c)} Ratio of the extinction-uncorrected to the
extinction-corrected $n_1 \phi(r)$ values for subsamples shown in panel
{\it a}.  {\it d)} Ratio of the extinction-uncorrected to the
extinction-corrected $n_1 \phi(r)$ values for subsamples shown in panel
{\it b}. }

\fig 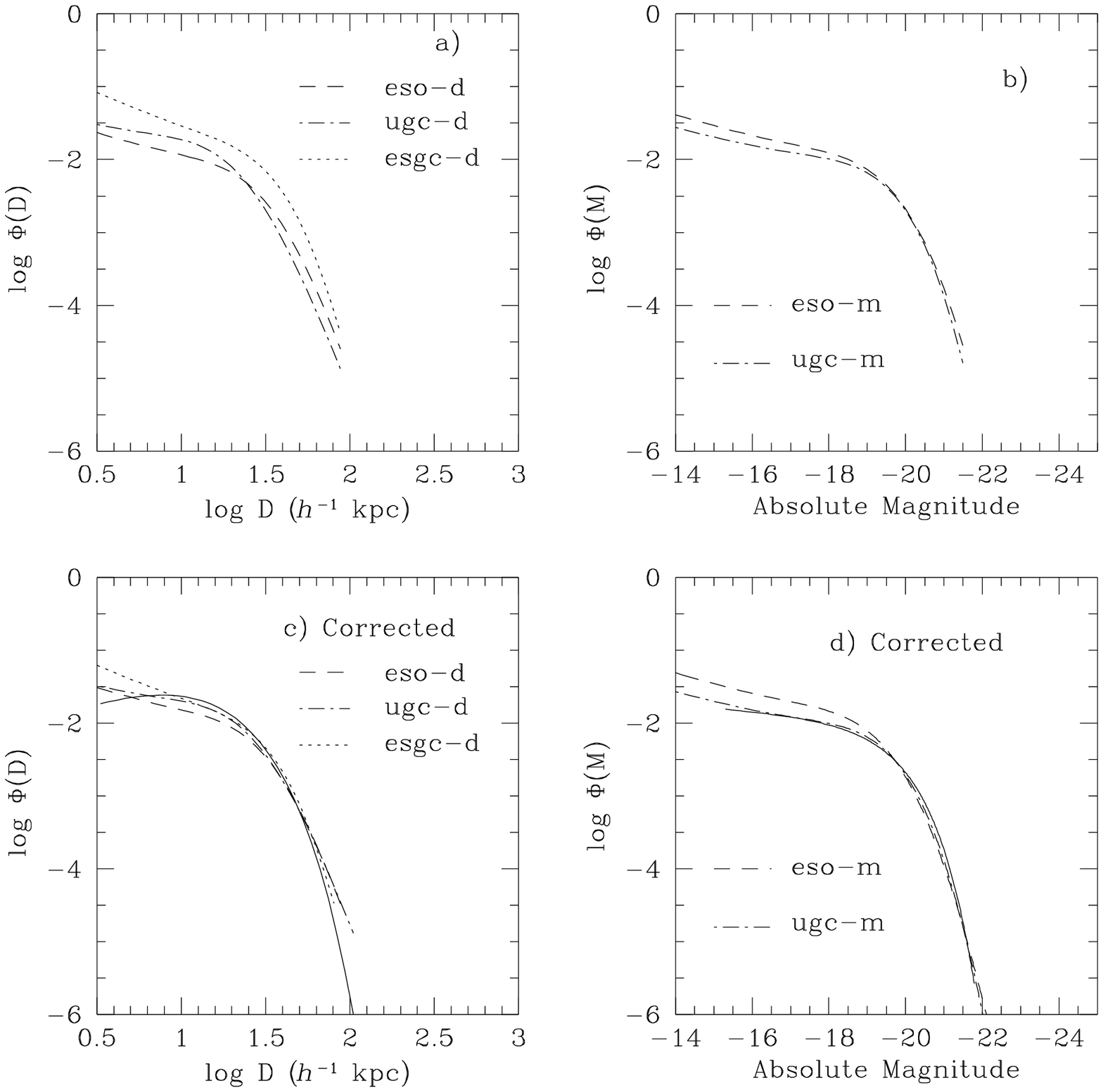, 5, 5, {9-  {\it a)} Derived diameter functions for the
ESO-d (dashed line), ESGC-d (dotted line) and UGC-d (dot-dashed line)
subsamples. The units are galaxies$\,$Mpc$^{-3}\,$(5 log
D/kpc)$^{-1}$.  {\it b)} Luminosity functions for the ESO-m (dashed)
and UGC-m (dot-dashed) subsamples. The units are
galaxies$\,$Mpc$^{-3}\,$mag$^{-1}$.  {\it c)} ESO-d, ESGC-d and UGC-d diameter
functions after taking into account the differences in mean density
and diameter scale between the three samples. The solid line is the
diameter function of CfA1 galaxies by Hudson \& Lynden-Bell (1991).
{\it d)} ESO-m and UGC-m luminosity functions after taking into
account the differences in mean density and magnitude scale between
the two samples. The solid line is the luminosity function of Loveday
\etal (1992), without any correction for Malmquist bias.}

\fig 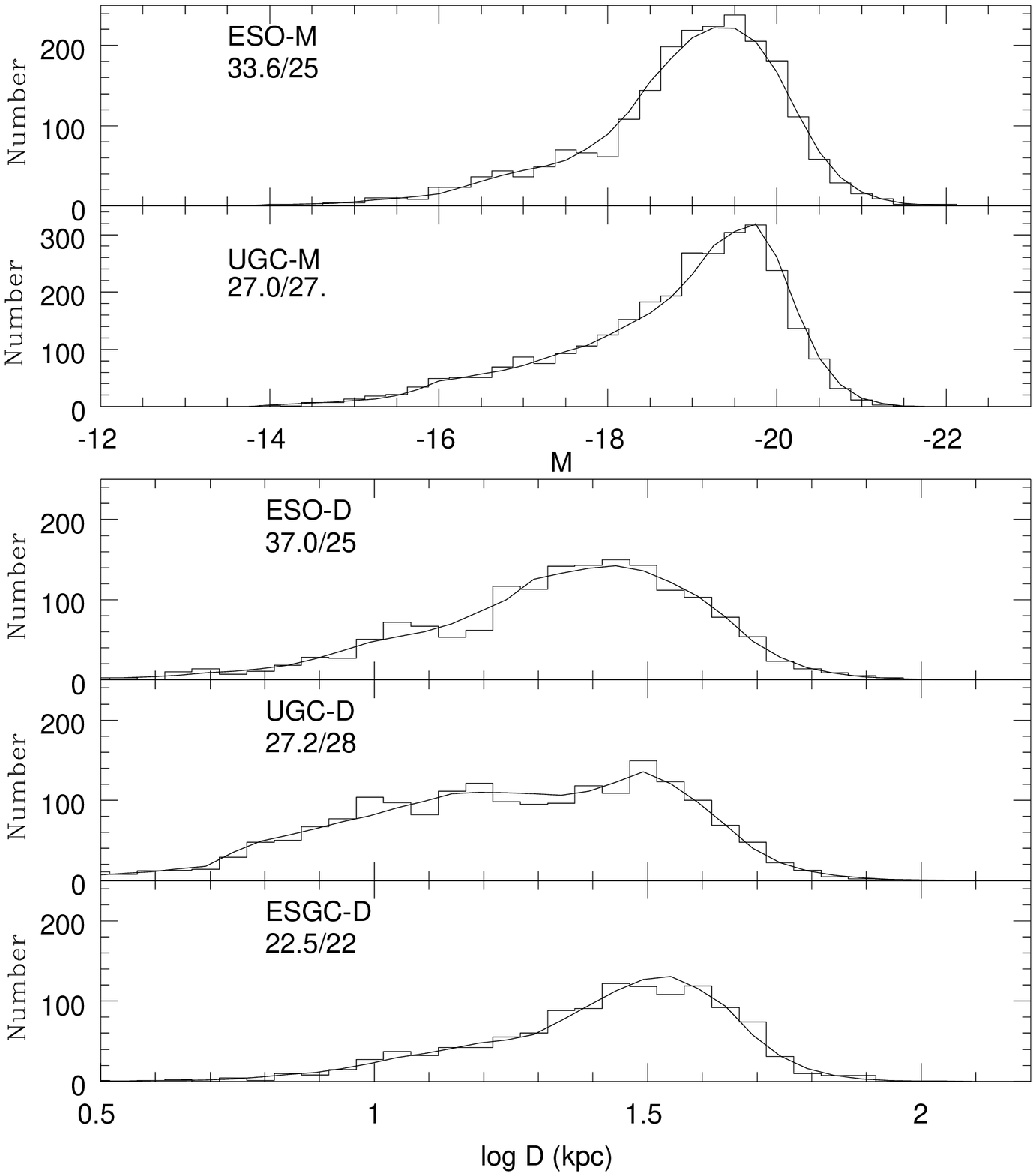, 5, 5, {10- Panel {\it a)} The histogram shows the luminosity
distribution of galaxies in the ESO-m subsample. The solid line
is the expected distribution from the best-fit solution to \lumf\
as given in Table~1. The goodness-of-fit parameter $\chi^{2}/{\rm dof}$ is
also shown.
Panel {\it b}:  as in {\it a} for the UGC-m subsample.
Panel {\it c}:  as in {\it a} for the diameter distribution of the ESO-d
subsample.
Panel {\it d}:  as in {\it c} for the UGC-d subsample.
Panel {\it e}:  as in {\it c} for the ESGC-d subsample.}

\fig 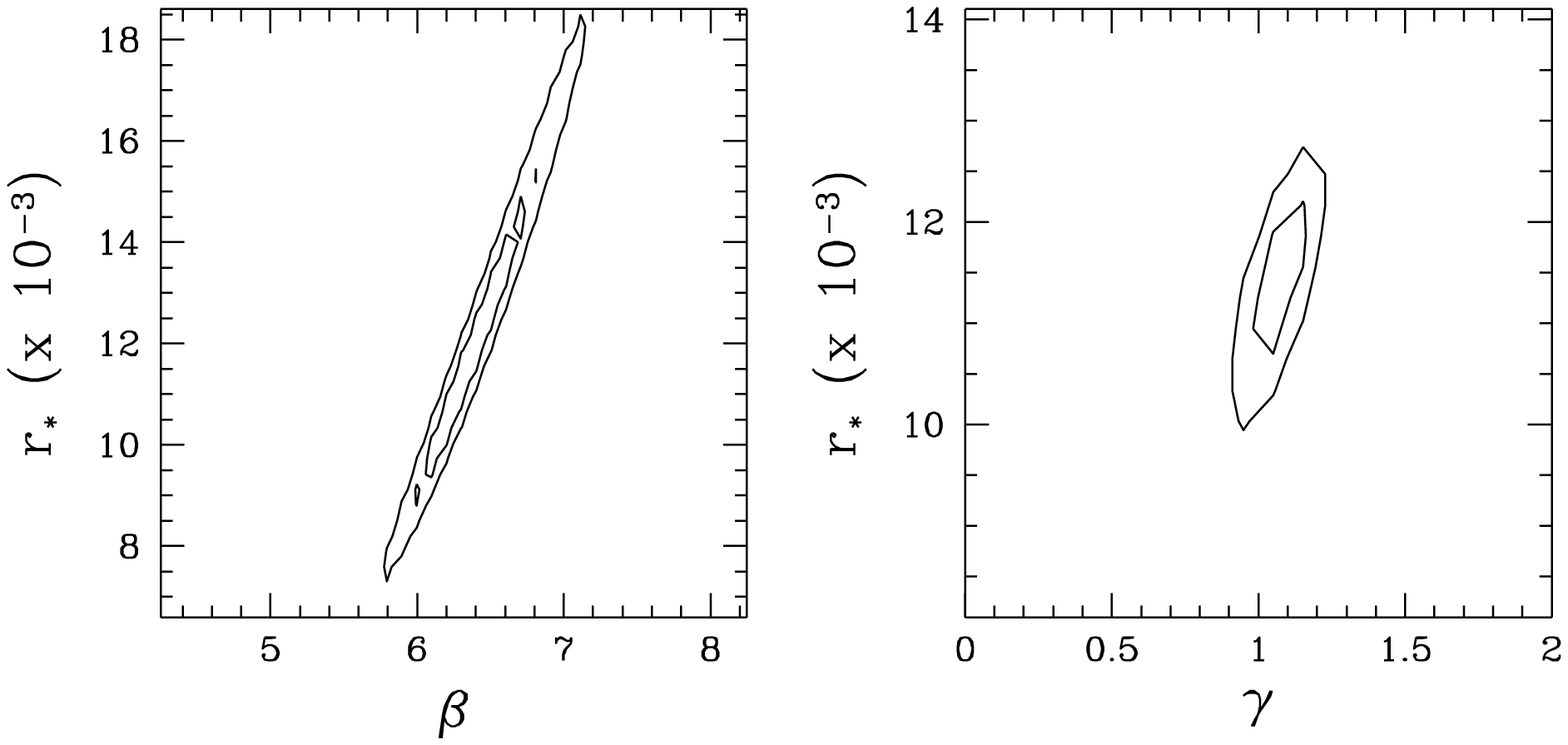, 5, 5, {11- 68\% and 90\% confidence level contours in the
likelihood function of ESO-m as projected on the $\beta-r_*$ (panel
{\it a}) and $\gamma - r_*$ (panel {\it b}) planes.}

%\fig multipanel.ps, 5, 5, {12- The density field of galaxies in the ORS, in 9
%%slices
%parallel to the principal planes in Supergalactic coordinates. The
%Supergalactic plane is in the center, with slices above and below it
%by 3000 \kms\ to the right and left, respectively. The upper panels
%are similar slices at constant SGX, and the lower are at constant
%SGY. Mean density is indicated by a heavy contour, and negative
%densities are indicated by dashed lines at $ \delta = -1/3$ and
%$-2/3$. The zone of avoidance is indicated by shading.}

\vfill\eject
\end